\documentclass{article}

\usepackage{authblk}
\usepackage{hyperref}

\usepackage{amsmath}

\usepackage{amsfonts}
\usepackage{amsmath}
\usepackage{amssymb}
\usepackage{framed}
\usepackage{braket}
\usepackage{float}

\usepackage{cmmib57}

\usepackage[dvips]{graphicx}
\usepackage{epsfig}
\usepackage{subcaption}

\usepackage[top=2cm, left=3.0cm, right=3.0cm, bottom=2cm]{geometry}

\newcommand{\ui}{{\mathrm{i}}}
\newcommand{\tph}{{t_{\mathrm{ph}}}}

\title{New approximation to compute the incoherent scattering function of harmonic lattices}
\author{V\'{\i}ctor Laliena\thanks{laliena@unizar.es} }
\author{Javier Campo\thanks{JAVIER.CAMPO@csic.es}}

\affil{Instituto de Ciencia de Materiales de Arag\'on (CSIC -- Universidad de Zaragoza) \\
and
Departamento de F\'{\i}sica de Materia Condensada, Universidad de Zaragoza \\
C/Pedro Cerbuna 12, E-50009 Zaragoza, Spain}

\date{July 19, 2018}

\begin{document}
\maketitle

\begin{abstract}
A new method to compute the incoherent scattering function of harmonic lattices is 
introduced. It is based in a saddle point approximation for each term of the phonon
expansion, and is simple enough to be used in practice. The method gives very accurate
results even for the tails of the scattering function, and is more accurate than the
usual gaussian approximation, which can be derived from this saddle point approximation
in the limit in which the order of the phonon expansion term becomes large. 
Numerical comparisons are provided using vanadium as a test case.
\end{abstract}

\section{Introduction}
\label{sec:intro}

The detailed knowledge of the neutron spectra at different points of space is important
for a fine analysis of neutron scattering experiments in condensed matter physics, as well
as in the design of neutron facilities, for technical reasons such as shielding, radiation 
protection, background estimation, etc.
The theoretical determination of the spectrum is complicated by the
fact that the scattering of neutrons of energies below a few eV is strongly affected by the 
target structure. In this case the scattering cross sections have a rich structure that depends
in a complicated way on the neutron energy and the target temperature. A simple and accurate 
determination of such detailed cross sections is important to estimate the scattering of neutrons
from components such as thermal neutron filters that remove the unwanted epithermal component of 
the spectrum~\cite{Brockhouse59}, or to provide corrections to scattering standards as 
vanadium~\cite{Mayers84}.

The scattering cross section of slow neutrons is completely determined by the target scattering 
function \cite{VanHove54}. For a single crystal material, the scattering function has an extremely
rich structure of peaks originated by the coherent scattering from single phonons. 
This structure is smoothed out in the case of multi-phonon scattering, and the multi-phonon part 
of the scattering function can be approximately obtained in the so called incoherent approximation, 
in which coherence effects are neglected \cite{Placzek55}. It has been argued that for a 
polycristalline material the coherence effects can also be neglected in the single phonon
scattering, due to the blurring caused by the average over the microstructure, and the
whole scattering function can be obtained in the incoherent approximation. This is very convenient
since the incoherent part of the scattering function is much more easier to evaluate than the
coherent part. 
Indeed, for solids that can be described as an harmonic system, the scattering function is 
completely determined by the phonon density of states (DoS).

In spite of the enormous simplification introduced by the incoherent approximation, 
to compute the scattering function in practical cases, for instance to get the cross sections in 
Monte Carlo simulations, it is necessary to make further approximations. 
The stardard way of computing the multi-phonon part of the scattering function is the gaussian 
approximation \cite{Sjolander58,Schofield59,Lovesey84}, in which each term of the phonon expansion 
beyond the one-phonon term is obtained as a gaussian function with mean and standard deviation
that increase with th order of the multi-phonon term, and which are easily calculable from the
DoS. The gaussian approximation works generally very well, although inaccuracies have been noticed
in some ranges of momentum and energy transfer \cite{Gunn84}. For high enough momentum transfer
the scattering function can be accurately obtained from the leading term of a saddle point expansion 
in powers of the inverse momentum transfer \cite{Egelstaff62}. 
This is however very inaccurate at low momentum transfer.

Recently, the sensitivity of the total scattering cross section to the DoS \cite{Holmes16},
and the influence on the total scattering cross section of modification of the DoS by anharmonic
effects such as phonon broading \cite{Cai17} have been studied. These works use the gaussian
approximation, which is rather accurate for the total cross section but causes notably departures
from the exact result for the differential cross section, which determines the spatial distribution 
of the neutron spectrum.

The present work focuses on a more accurate computation of the scattering function, introducing a new
approximation that is still simple enough to be used in practice but is far more accurate than the 
gauusian approximation for the lowest order terms of the phonon expansion, thus providing more accurate 
differential cross sections for some energy ranges.
It is based on a saddle point expansion for each term of the phonon expansion. 
Within the present approximation, the gaussian approximation is attained asymptotically as the order of 
phonon expansion increases. Thus, it can be viewed as a further approximation to the saddle point expansion.
The approximation developed in Ref.~\cite{Egelstaff62} is also based on a saddle point
expansion, which however is very different from the saddle point expansion introduced here.
To distinguish them we call the former the saddle point approximation for the full scattering 
function (SPFS) and the latter the saddle point approximation for the phonon expansion (SPPE).

Let us remark that the DoS is an input for the incoherent approximation of the scattering function, and thus 
the results of this paper are independent of the way the DoS is obtained (\textit{ab-initio} computations
or experimentally measured).

Besides more accurate computations of the scattering cross sections in numerical simulations,
a more accurate determination of the incoherent scattering function can be used for interpreting experimental
results in condensed matter physics via the refinement of the analysis of neutron scattering experiments.
It can also be useful to obtain the DoS by fitting experimental results to a properly parametrized DoS.
 
The paper is organized as follows. In Sec.~\ref{sec:sf} the main properties of the scattering function
for harmonic lattices are reviewed in order to set clearly the problem and to introduce the notation.
In Sec.~\ref{sec:sp} the saddle point approximation for the full scattering function
is briefly described for comparison with the saddle point approximation for the phonon expansion,
which is developed in Sec.~\ref{sec:phe}. The gaussian approximation is derived 
from the saddle point approximation for the phonon expansion in Sec.~\ref{sec:gauss}. 
A convenient representation of the scattering 
function, similar to that proposed in \cite{Cuello97}, is introduced in Sec.~\ref{sec:rep}.
Section~\ref{sec:num} is devoted to describe the numerical computations performed to compare the 
different approximations considered in this work.
The paper ends with a summary of the conclusions. Some mathematical technical details are given in
the appendix.

\section{The scattering function for harmonic lattices}
\label{sec:sf}

The incoherent differential scattering cross section for an incident neutron of wave vector $\vec{k}$ 
and energy $E$
scattered into a neutron of wave vector $\vec{k}^\prime$ and energy $E^\prime$
can be written in terms of the target scattering function \cite{VanHove54}
\begin{equation}
S(\vec{q},\omega) = \int_{-\infty}^\infty\frac{dt}{2\pi\hbar}\mathrm{e}^{-\ui\omega t}
\left\langle \mathrm{e}^{\ui\vec{q}\cdot\vec{r}(0)}\mathrm{e}^{-\ui\vec{q}\cdot\vec{r}(t)}\right\rangle
\label{eq:Sdef}
\end{equation}
as
\begin{equation}
\frac{d^2\sigma}{d\Omega dE^\prime} = N\frac{k^\prime}{k}\frac{\sigma_{\mathrm{inc}}}{4\pi}S(\vec{q},\omega),
\label{eq:XS}
\end{equation}
where $N$ is the number of scattering centers in the target, 
$\vec{q}=\vec{k}-\vec{k}^\prime$ is the scattering vector, and $\omega=(E-E^\prime)/\hbar$.
For simplicity, we assume that all particles in the target are equivalent, so that $\vec{r}(t)$ represents
the Heisenberg position operator for one particle. This means that the discussion is limited to Bravais 
lattices.
The brackets in Eq.~(\ref{eq:Sdef}) denote the thermal average, which,
for a target of particles that interact through isotropic harmonic forces, can be written as
\begin{equation}
\left\langle \mathrm{e}^{\ui\vec{q}\cdot\vec{r}(0)}\mathrm{e}^{-\ui\vec{q}\cdot\vec{r}(t)}\right\rangle
= \exp\left\{q^2[G(t)-G(0)]\right\},
\end{equation}
where
\begin{equation}
G(t) = \langle \vec{r}(t)\cdot\vec{r}(0) \rangle
\end{equation}
is the self-correlation function,
which in its turn can be obtained from the density of states (DoS) of the target, $Z(\omega)$, 
as\footnote{See Ref.~\protect\cite{Lovesey84}, pags. 162-163.}
\begin{equation}
G(t) = \frac{\hbar}{2M}\int_{-\infty}^\infty d\omega \frac{Z(\omega)}{\omega}n(\omega)\mathrm{e}^{-\ui\omega t},
\end{equation}
where $M$ is the particle mass and $n(\omega)$ is the Bose occupancy number
\begin{equation}
n(\omega) = \frac{1}{\exp(\hbar\omega/k_\mathrm{B}T)-1},
\end{equation}
at temperature $T$, with $k_\mathrm{B}$ being the Boltzman constant.

The DoS has a cut-off, $\omega_{\mathrm{m}}$, so that it vanishes for $\omega>\omega_{\mathrm{m}}$.
In what follows it is convenient to work with the following dimensionless quantities, using $\omega_{m}$ as a
characteristic parameter: the dimesionless time $s=\omega_{\mathrm{m}}t$; the dimensionless frequency
(or energy) $u=\omega/\omega_{m}$; the dimensionless momentum transfer,
$Q^2=\hbar q^2/2M\omega_{\mathrm{m}}$; the dimensionless DoS, 
\begin{equation}
Z(u)=\omega_{\mathrm{m}}Z(\omega_{\mathrm{m}}u);
\end{equation}
and the dimensionless self-correlation function, 
\begin{equation}
\gamma(s) = \omega_{\mathrm{m}}G(s/\omega_{\mathrm{m}}) = 
\int_{-\infty}^\infty du \frac{Z(u)}{u} n(u\omega_{\mathrm{m}}) \mathrm{e}^{-\ui us}.
\label{eq:gdef}
\end{equation}
The cross section reads
\begin{equation}
\frac{d^2\sigma}{d\Omega dE^\prime} = N\frac{\sigma_{\mathrm{inc}}}{4\pi}
\frac{1}{E_{\mathrm{m}}} \sqrt{1+u\frac{E_{\mathrm{m}}}{E}} S(Q^2,u),
\end{equation}
where $E_{\mathrm{m}}=\hbar\omega_{\mathrm{m}}$ and the dimensionless scattering function, for which
we used the same symbol as for the dimensionful function, to avoid symbol proliferation,
is given by
\begin{equation}
S(Q^2,u) = \int_{-\infty}^\infty \frac{ds}{2\pi} \exp\{-\ui us+Q^2[\gamma(s)-\gamma(0)]\}.
\label{eq:S}
\end{equation}
We will use the convention that the DoS is an even function of $u$, by defining $Z(u)=Z(-u)$ for $u<0$.

\section{The saddle point approximation for the full scattering function}
\label{sec:sp}

For high momentum transfer, $Q^2\rightarrow\infty$, the scattering function can be evaluated by the
saddle point method \cite{Egelstaff62}. In that limit the energy transfer is also very large, 
$u\sim Q^2$, so that we introduce the variable $\xi=u/Q^2$. The integration contour in
Eq.~(\ref{eq:S}) is deformed in the complex plane so that it passes through a saddle point
of $-\ui\xi s + \gamma(s)$, determined by the equation
\begin{equation}
\gamma^{\,\prime}(s) = \ui\xi,
\label{eq:sp}
\end{equation}
where the prime stands for the derivatives with respect to $s$. The solution to the above equation
is a purely imaginary number, $s=\ui t_{\mathrm{sp}}(\xi)$.
The scattering function is asymptotically equal to \cite{Egelstaff62}
\begin{equation}
S_{\mathrm{SP}}(Q^2,u) = \frac{1}{\sqrt{2\pi Q^2[-\gamma^{\prime\prime}(\ui t_{\mathrm{sp}})]}}
\exp\left\{ut_{\mathrm{sp}}+Q^2[\gamma(\ui t_{\mathrm{sp}})-\gamma(0)]\right\}
\label{eq:Ssp}
\end{equation}
where $t_{\mathrm{sp}}(\xi)$ is evaluated at $\xi=u/Q^2$.
The properties of the saddle point solution have been thoroughly analyzed in Ref.~\cite{Gunn84}.

\section{The saddle point expansion for the phonon expansion}
\label{sec:phe}

The saddle point evaluation of the full scattering function fails at low momentum transfer, when 
$Q^2\lesssim 1$. For low enough $Q^2$ it is used the phonon expansion
\begin{equation}
S(Q^2,u) = \exp(-Q^2\gamma_0)\sum_{p=0}^\infty\frac{1}{p!}(Q^2\gamma_0)^pF_p(u)
\label{eq:phe}
\end{equation}
where $\gamma_0 = \gamma(0)$ and
\begin{equation}
F_p(u) = \int_{-\infty}^\infty\frac{ds}{2\pi}\left[\frac{\gamma(s)}{\gamma_0}\right]^p \mathrm{e}^{-\ui us}.
\label{eq:Fp}
\end{equation}
The functions $F_p(u)$ are normalized to unity,
\begin{equation}
\int_{-\infty}^\infty F_p(u)du = 1,
\end{equation}
and satisfy the following recursion relation 
\begin{equation}
F_p(u) = \int_{-\infty}^\infty F_1(u-u^\prime)F_{p-1}(u^\prime)du^\prime
\label{eq:rr}
\end{equation}
for $p\geq 1$, with
\begin{eqnarray}
F_0(u) &=& \delta(u), \\
F_1(u) &=& -\frac{Z(u)}{u\gamma_0} n(-u\omega_{\mathrm{m}}).
\end{eqnarray}
Notice that, due to the $Z(u)$ factor, $F_1(u)$ vanishes for $|u|\geq 1$. The recursion relation~(\ref{eq:rr})
then implies that $F_p(u)$ vanishes for $|u|\geq p$.

The phonon expansion is very useful at low momentum transfer, where a few terms suffices to 
get $S(Q^2,u)$ with good accuracy. There is a regime, however, where the convergence of the phonon expansion 
is slow and the saddle point approximation $S_{\mathrm{SP}}(Q^2,u)$ is an inaccurate representation of the 
scattering function. 
In this case many terms of the phonon series have to be added and therefore it is convenient to have a good
way of evaluating $F_p(u)$. The gaussian approximation, which will be briefly reviewed in the next section,
provides a simple and in most cases accurate representation of $F_p(u)$ by a gaussian function. Here we develop
a method for computing $F_p(u)$ that is more accurate than the gaussian method and simple enough to be used in
practice.

The method is based again in a saddle point expansion, this time for large $p$. To this end, let us rewrite 
Eq.~(\ref{eq:Fp}) as
\begin{equation}
F_p(u) = \int_{-\infty}^\infty\frac{ds}{2\pi}
\exp\left\{p\left[\log\frac{\gamma(s)}{\gamma_0} -\ui\frac{u}{p}s \right]\right\}.
\label{eq:Fp2}
\end{equation}
The argument of the exponential in the integral is an analytic function of $s$ in a neighbourhood of $s=0$ and
thus the contour can be deformed in the complex plane to pass through the saddle point, which is the solution
of the equation
\begin{equation}
\frac{\gamma^{\,\prime}(s)}{\gamma(s)} = \ui\xi,
\label{eq:spph}
\end{equation}
where $\xi=u/p$. The solution of this equation is a purely imaginary number denoted 
by $\ui\tph (\xi)$. 

The argument of the exponential in the integrand of Eq.~(\ref{eq:Fp2}) can be expanded in Taylor series around
the saddle point. The linear term in \mbox{$s-\ui\tph (\xi)$} vanishes on account of the saddle point 
equation,~(\ref{eq:spph}), and we have
\begin{equation}
\log\frac{\gamma(s)}{\gamma_0} -\ui\frac{u}{p}s =
\log\frac{\gamma(\ui\tph )}{\gamma_0} + \frac{u}{p}\tph 
+ \frac{1}{2}
\left[\frac{\gamma^{\,\prime\prime}(\ui\tph )}{\gamma(\ui\tph )}+\frac{u^2}{p^2}\right]
\left(s-\ui\tph\right)^2+\ldots
\label{eq:taylor}
\end{equation}
The integration contour in the neighborhood of the saddle point can be chosen parallel to the real axis:
\mbox{$s-\ui\tph (\xi)=r$}. The main contribution to the integral comes from this neighborhood,
so that we plug the expansion~(\ref{eq:taylor}) into Eq.~(\ref{eq:Fp2}) and perform the gaussian integration in $r$.
The result is that as $p\rightarrow\infty$ the function $F_p(u)$ is asymptotic to
\begin{equation}
F_p^{\mathrm{(sp)}}(u) =  
\left[-2\pi p \left( \frac{ \gamma^{\,\prime\prime}(\ui\tph )} {\gamma(\ui\tph )}
+\frac{u^2}{p^2}\right)\right]^{-1/2} 
\left[\frac{ \gamma(\ui\tph ) }{\gamma_0} \right]^p \exp(u\tph ),
\label{eq:Fpsp}
\end{equation}
where $\tph $ is a function of $\xi=u/p$. Notice that the fact that $p\rightarrow\infty$ does not imply
$\xi\rightarrow 0$, since the typical values of $u$ grow with $p$.
The saddle point approximation to $F_p(u)$ is extremely accurate for $p>2$, and rather good for $p=2$,
as we will see. Furthermore, it is remarkable that it vanishes for $|u|\geq p$, as $F_p(u)$ does.
This is proven in the appendix.
Thus, this saddle point approximation reproduces accurately even the tails of $F_p(u)$.

\begin{figure}[t!]
\centering
\includegraphics[width=0.32\linewidth,angle=0]{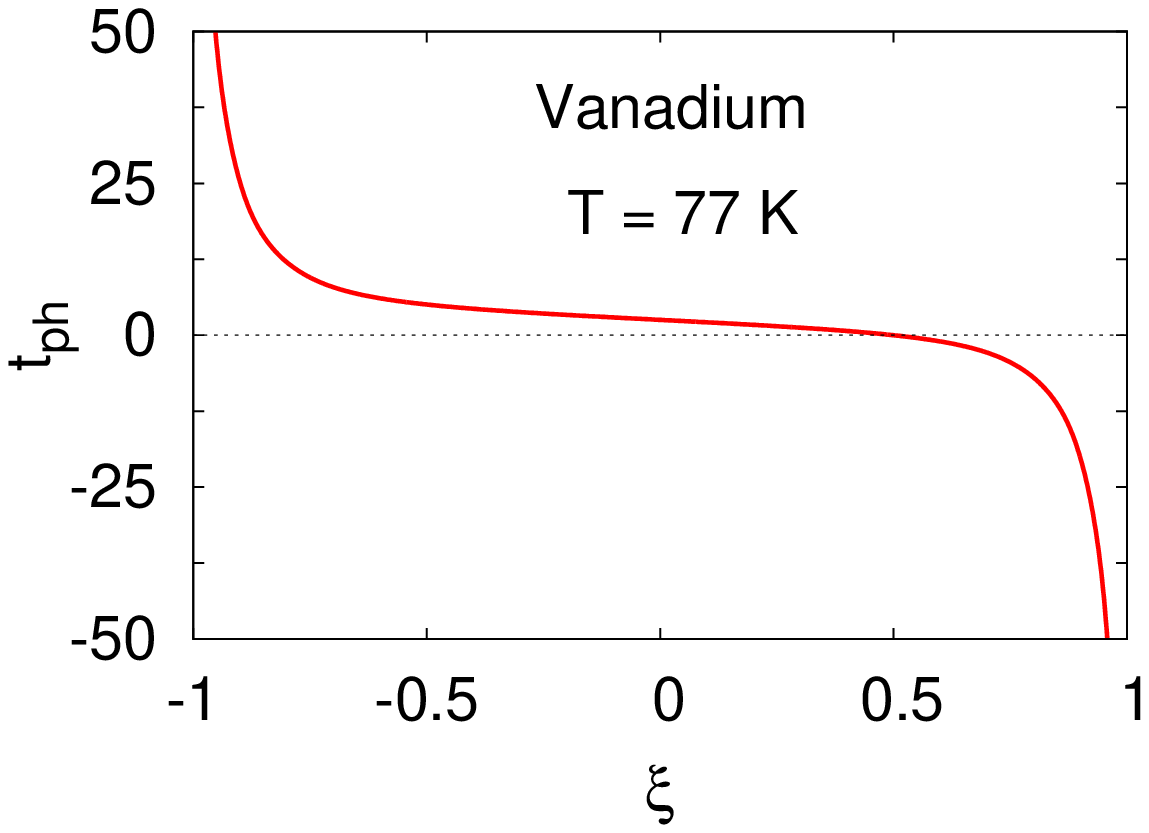}
\includegraphics[width=0.32\linewidth,angle=0]{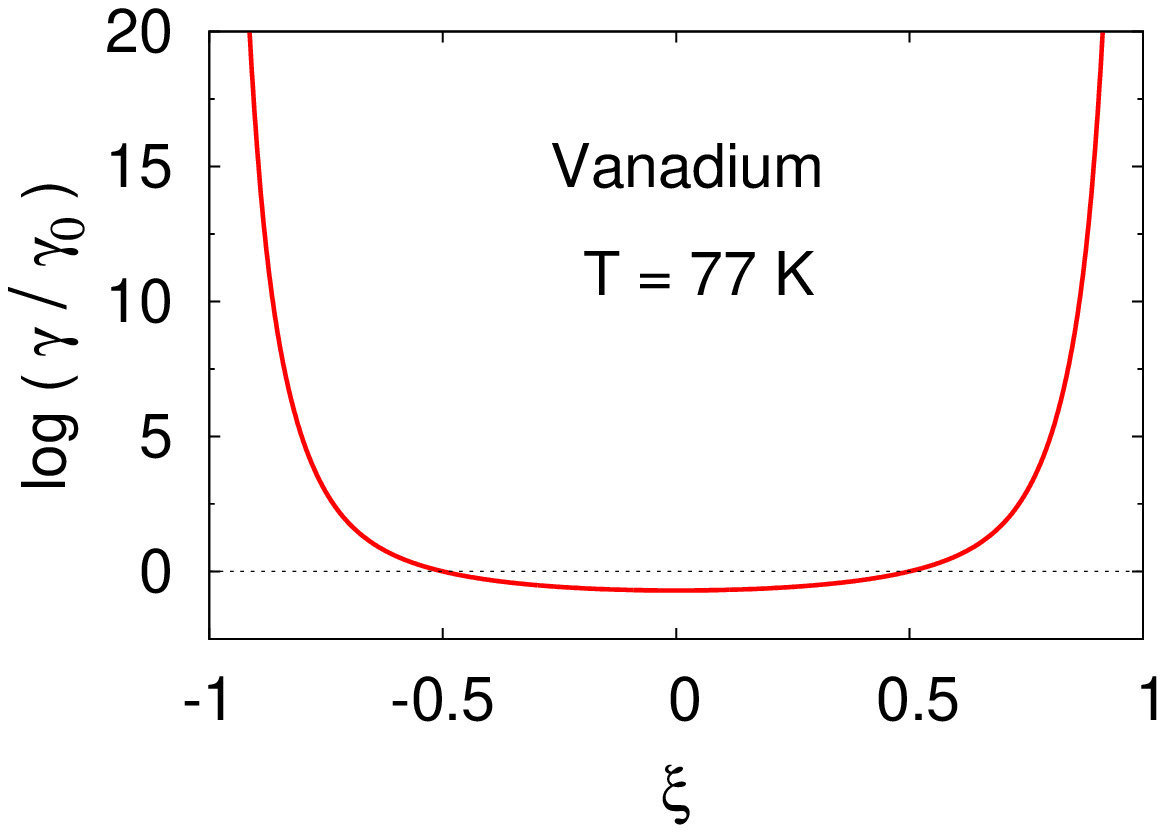}
\includegraphics[width=0.32\linewidth,angle=0]{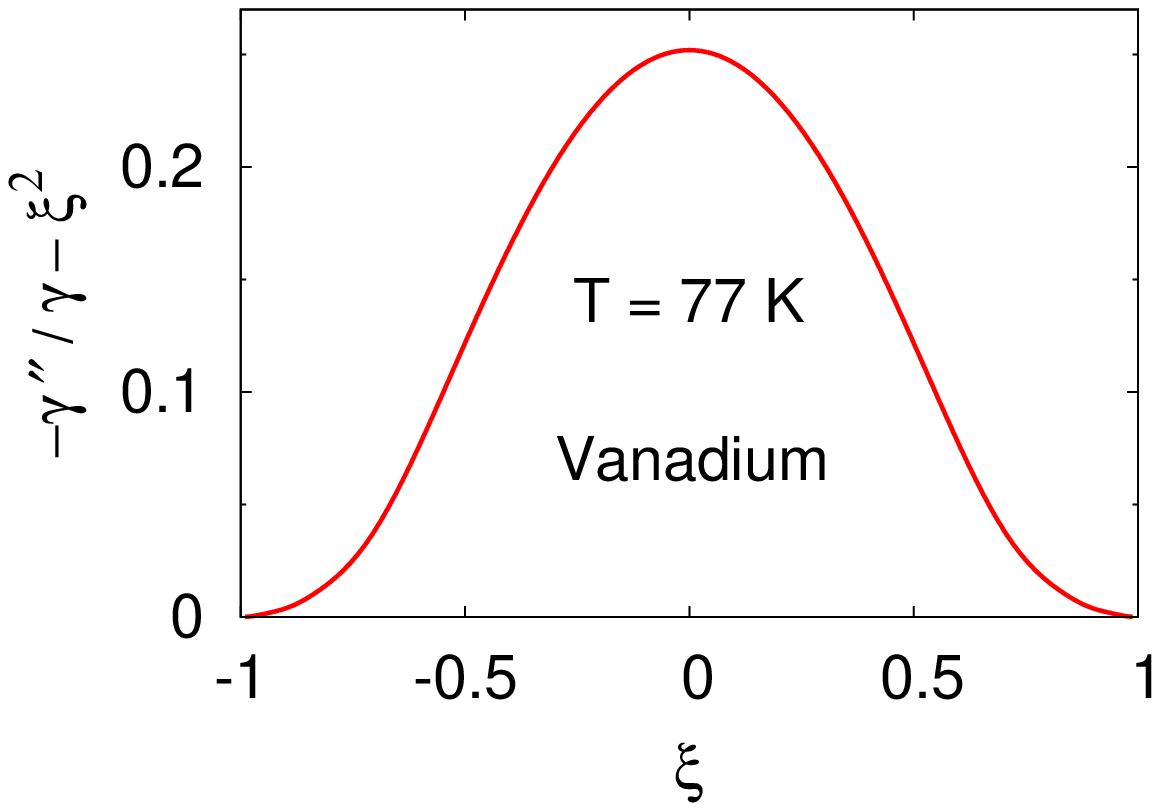}
\caption{The solution of the saddle point equation~(\ref{eq:spph}) and the functions
entering Eq.~(\ref{eq:Fpsp}) as a function of $\xi$ for vanadium at \mbox{77 K}.
The argument of the functions $\gamma$ and $\gamma^{\,\prime\prime}$ is $\ui\tph(\xi)$.
\label{fig:tph}}
\end{figure}

Fig.~\ref{fig:tph} (left) displays $\tph$ as a function of $\xi$ for vanadium at \mbox{$T=77$ K}, computed
with the measured DoS published by Sears \textit{et al.}~\cite{Sears95}. The middle and right panels
display the functions entering~(\ref{eq:Fpsp}).

\section{The gaussian approximation}
\label{sec:gauss}

The gaussian approximation \cite{Sjolander58,Schofield59,Lovesey84} assumes that in multi-phonon effects
the integral in Eq.~(\ref{eq:Fp2}) gets the main contribution form the neighbourhood of $s=0$.
The main contribution, however, comes from the imaginary saddle point, $\ui\tph$, which vanishes 
at $\xi=1/\gamma_0$,
and diverges for $\xi\rightarrow\pm 1$. Thus, the gaussian approximation can only be accurate for
$u$ in the neighborhood of $p/\gamma_0$. It is proven in the appendix that $\tph$ tends to zero as
$p\rightarrow\infty$ for fixed $u$. Then, the gaussian approximation improves by increasing $p$, and
becomes essentially exact as $p\rightarrow\infty$. This is not surprising, since the gaussian
approximation is connected to the central limit theorem, which is asymptotically valid for large $p$
\cite{Sjolander58,Schofield59}. 

In the gaussian approximation the integral of Eq.~(\ref{eq:Fp2}) is evaluated by means of an 
expansion in powers of $s$ up to the second order:
\begin{equation}
\log\frac{\gamma(s)}{\gamma_0} = \frac{\ui}{\gamma_0} s - \frac{1}{2}\Delta^2s^2+\ldots
\end{equation}
where 
\begin{equation}
\Delta^2 = -\frac{\gamma^{\,\prime\prime}(0)}{\gamma_0}-\frac{1}{\gamma_0^2}
\label{eq:Delta2}
\end{equation}
and we used the fact that $\gamma^{\,\prime}(0)=\ui$. Notice that, as proven in the appendix,
$\Delta^2>0$.
The integral in $s$ is therefore gaussian an can be readily 
performed. Thus we get the gaussian approximation for $F_p^{\mathrm{(sp)}}(u)$:
\begin{equation}
F_p^{\mathrm{(g)}}(u) =  
\frac{1}{\sqrt{2\pi p\Delta^2}}
\exp\left[-\frac{1}{2p\Delta^2}\left(u-\frac{p}{\gamma_0}\right)^2\right].
\label{eq:Fpg}
\end{equation}
The result is therefore a gaussian form for $F_p^{\mathrm{(sp)}}(u)$ with mean $p/\gamma_0$ and 
standard deviation $\sqrt{p\Delta^2}$. Notice that both the mean and the standard deviation increase 
with $p$. If the rescaled variable $\xi=u/p$ is used, the mean is $1/\gamma_0$, 
independent of $p$, and the gaussian is sharply peaked as $p$ increases, since the standard 
deviation is $\sqrt{\Delta^2/p}$.

\section{Representation of the scattering function}
\label{sec:rep}

For practical purposes, the scattering function can be represented by taking the exact first
$n_{\mathrm{ph}}$ terms of the phonon expansion, with $n_{\mathrm{ph}}>0$, as
\begin{equation}
S(Q^2,u) = \mathrm{e}^{-Q^2\gamma_0}\sum_{p=0}^{n_{\mathrm{ph}}}\frac{1}{p!}(Q^2\gamma_0)^pF_p(u)
+ W_{n_{\mathrm{ph}}+1}(Q^2) S_{\mathrm{MP}}(Q^2,u,n_{\mathrm{ph}}+1),
\label{eq:Sprac}
\end{equation}
where $S_{\mathrm{MP}}(Q^2,u,n_{\mathrm{ph}}+1)$ represents the contribution of the multi-phonon terms
with $p\geq n_{\mathrm{ph}}$, and the factor $W_{n_{\mathrm{ph}}+1}(Q^2)$ ensures the proper weight to
the multi-phonon contribution:
\begin{equation}
W_{n_{\mathrm{ph}}+1}(Q^2) = \frac{1-\mathrm{e}^{-Q^2\gamma_0}\sum_{p=0}^{n_{\mathrm{ph}}}\frac{1}{p!}[Q^2\gamma_0]^p}
{\int S_{\mathrm{MP}}(Q^2,u,n_{\mathrm{ph}}+1)du},
\label{eq:R}
\end{equation}
so that $\int S(Q^2,u) du = 1$. For $S_{\mathrm{MP}}$ we may use either the SPPE or the
gaussian approximation, so that
\begin{equation}
S_{\mathrm{MP}}(Q^2,u,n_{\mathrm{ph}}+1) = 
\exp(-Q^2\gamma_0)\sum_{p=n_{\mathrm{ph}+1}}^{n_{\mathrm{max}}}\frac{1}{p!}[Q^2\gamma_0]^pF_p^{(a)}(u),
\label{eq:SMPprac}
\end{equation}
where $n_{\mathrm{max}}$ is the maximum number of multi-phonon terms included in $S_{\mathrm{MP}}$ and
the superscript $(a)$ stands for (sp), in the case of the SPPE, or for (g), if the gaussian 
approximation is used. 

The SPFS, Eq.~(\ref{eq:Ssp}), can also be used for the multi-phonon part of the scattering function,
but this is only accurate if $Q^2$ is large enough. For vanadium at \mbox{294 K} it is rather accurate 
for $Q^2\gtrsim 1$, but at \mbox{77 K} it is necessary $Q^2\gtrsim 3$.

Cuello \textit{et al.} \cite{Cuello97} 
proposed to use $n_{\mathrm{ph}}=3$ in a representation of the scattering function similar 
to~(\ref{eq:Sprac}), with a different approximation for the multi-phonon contribution.

\begin{figure}[t!]
\centering
\includegraphics[width=0.32\linewidth,angle=0]{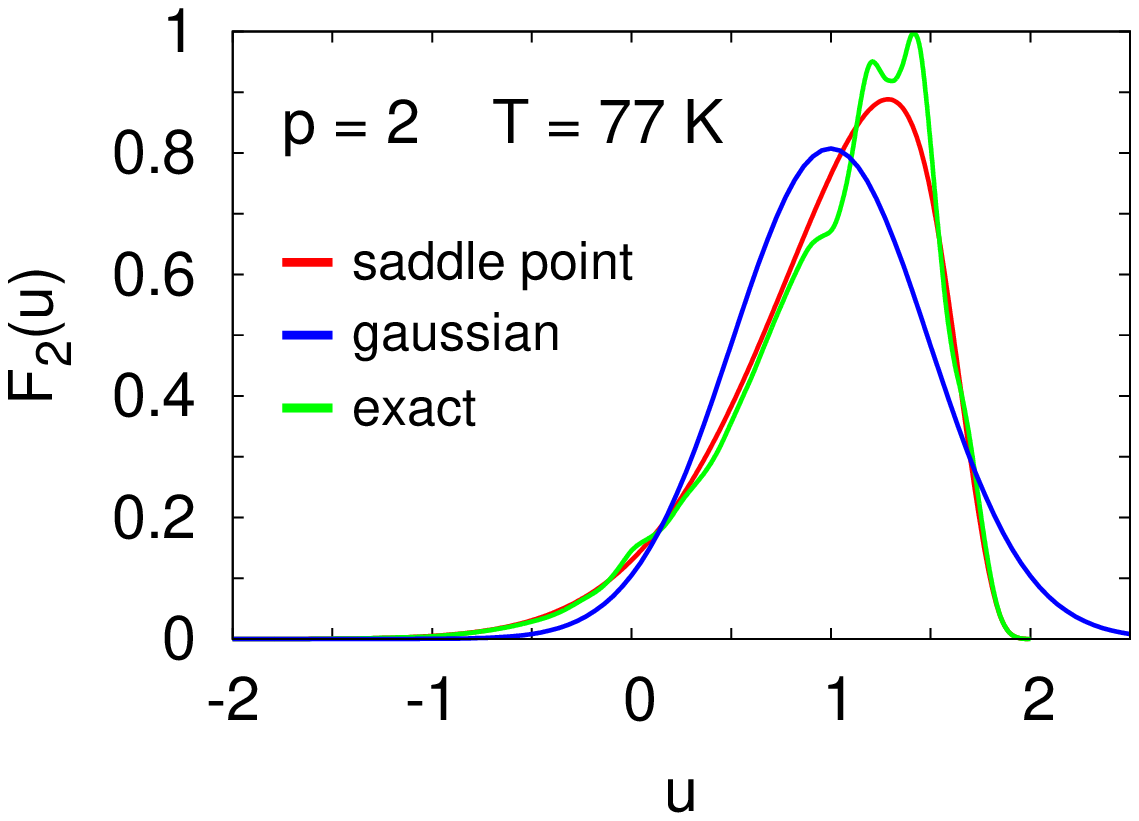}
\includegraphics[width=0.32\linewidth,angle=0]{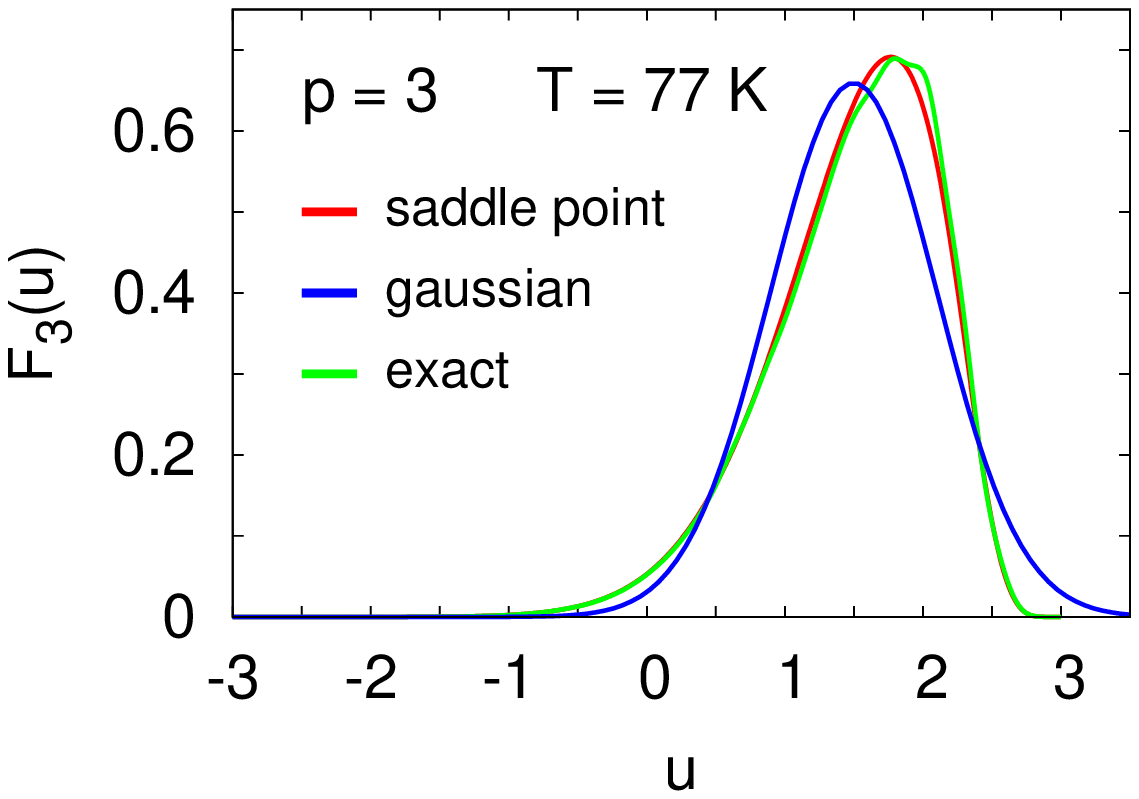}
\includegraphics[width=0.32\linewidth,angle=0]{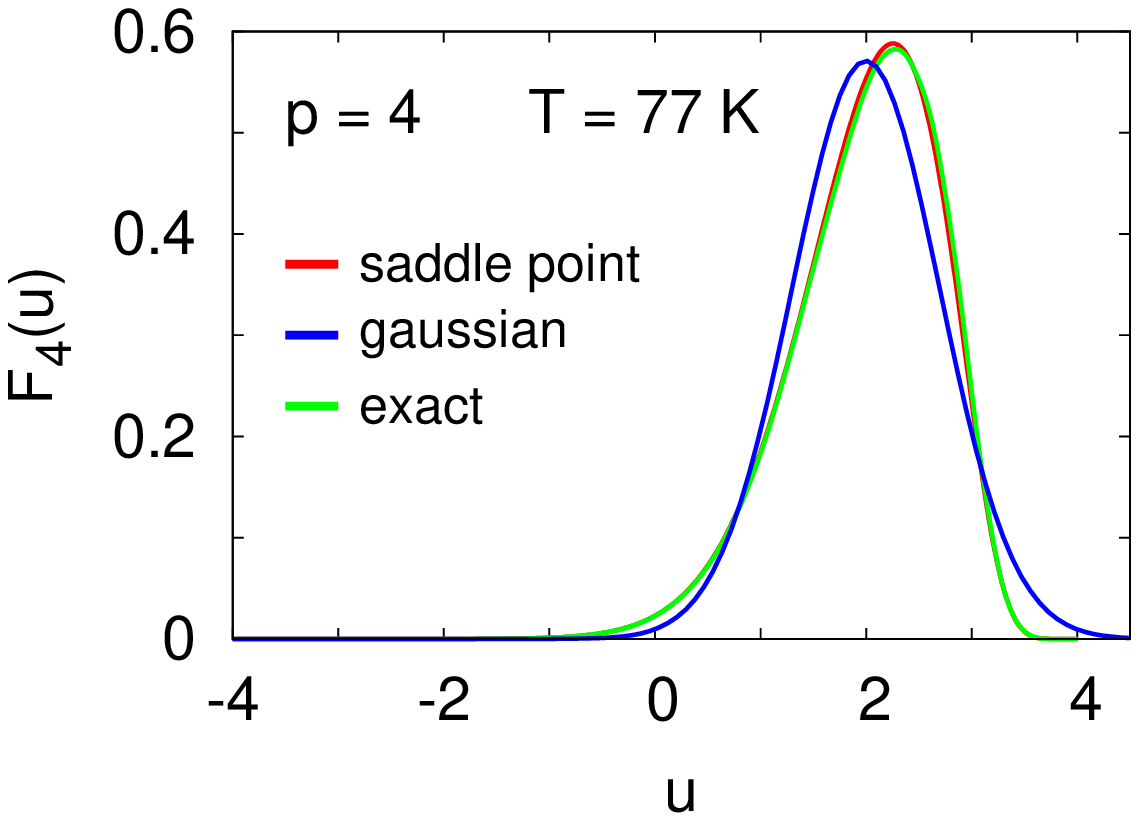}

\includegraphics[width=0.32\linewidth,angle=0]{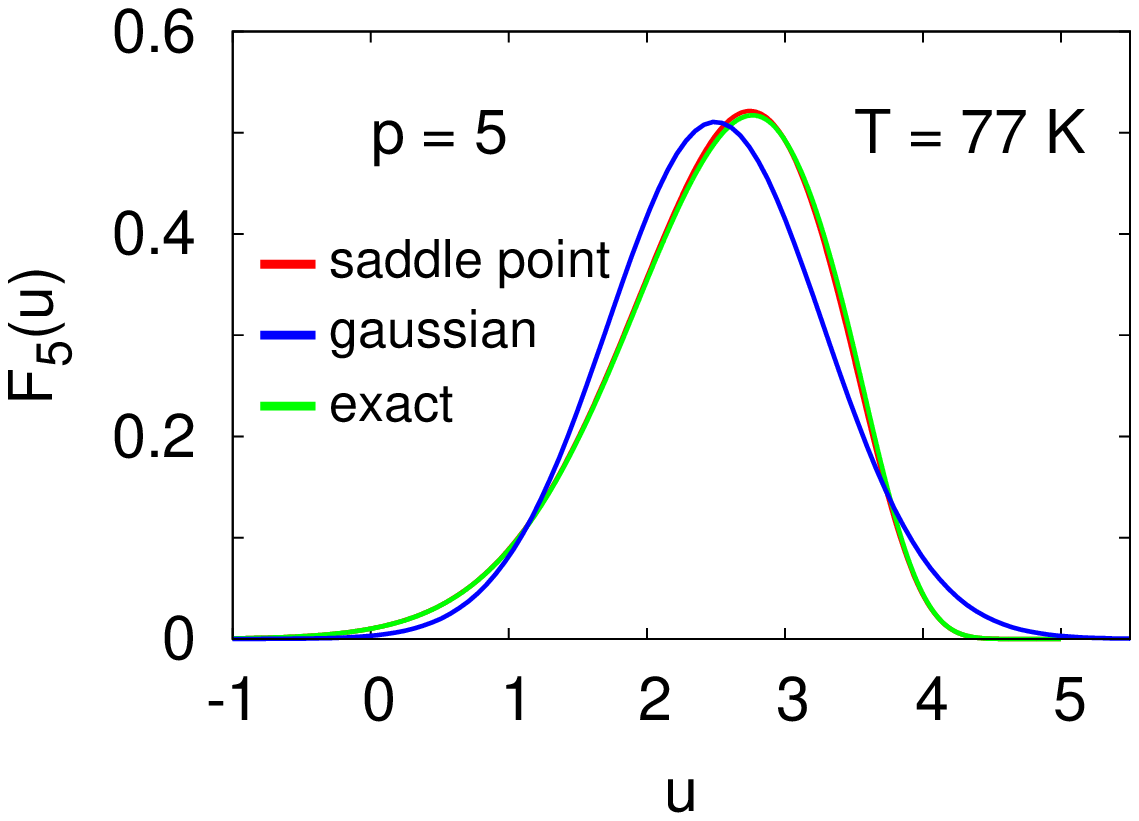}
\includegraphics[width=0.32\linewidth,angle=0]{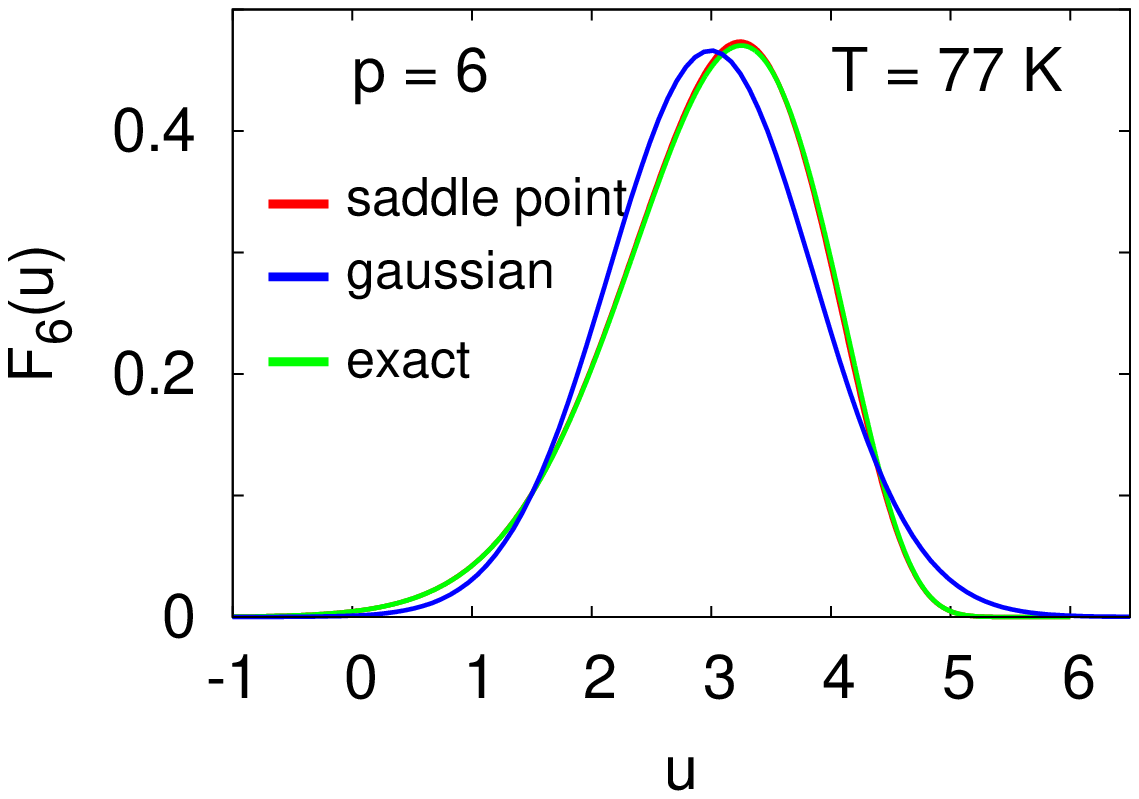}
\includegraphics[width=0.32\linewidth,angle=0]{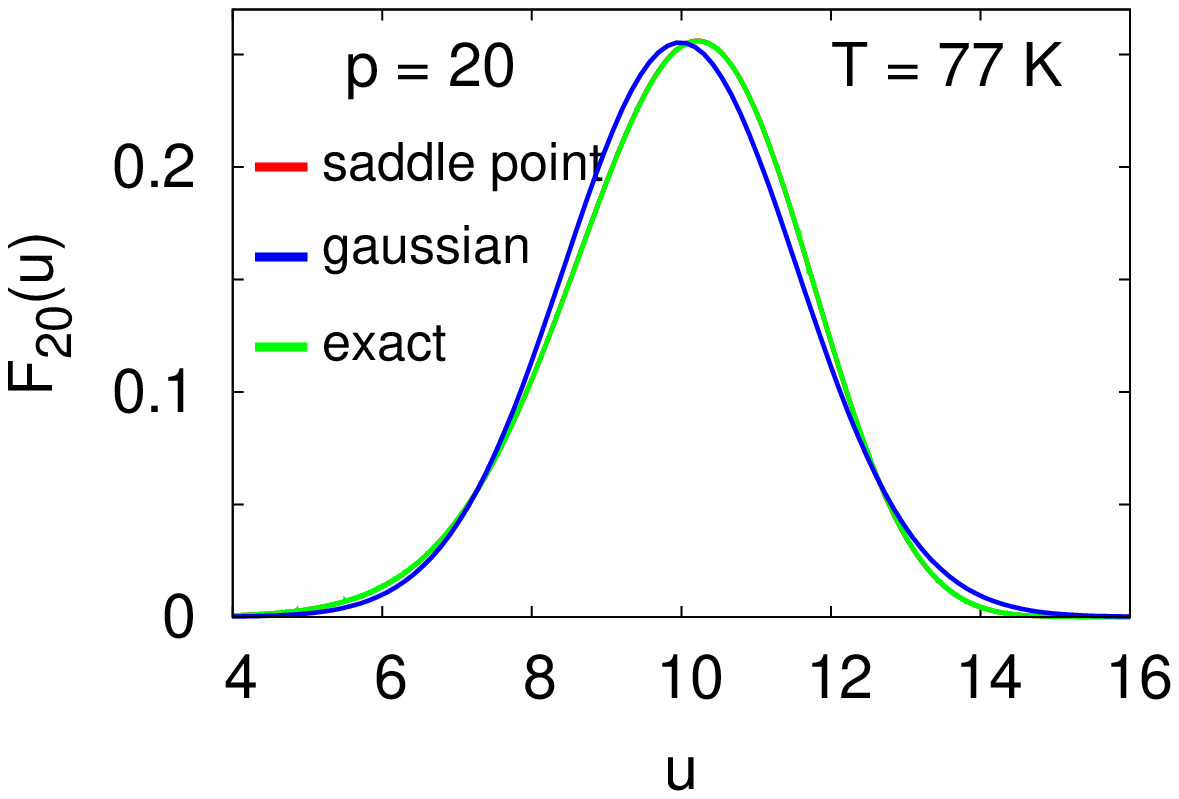}
\caption{The function $F_p(u)$ for vanadium at \mbox{77 K} computed in three ways: 
Eq.~(\ref{eq:Fpsp}), saddle point approximation (red);
Eq.~(\ref{eq:Fpg}), gaussian approximation (blue); 
and exact, computed by numerical integration (green).
\label{fig:Fps77}}
\end{figure}

\begin{figure}[t!]
\centering
\includegraphics[width=0.32\linewidth,angle=0]{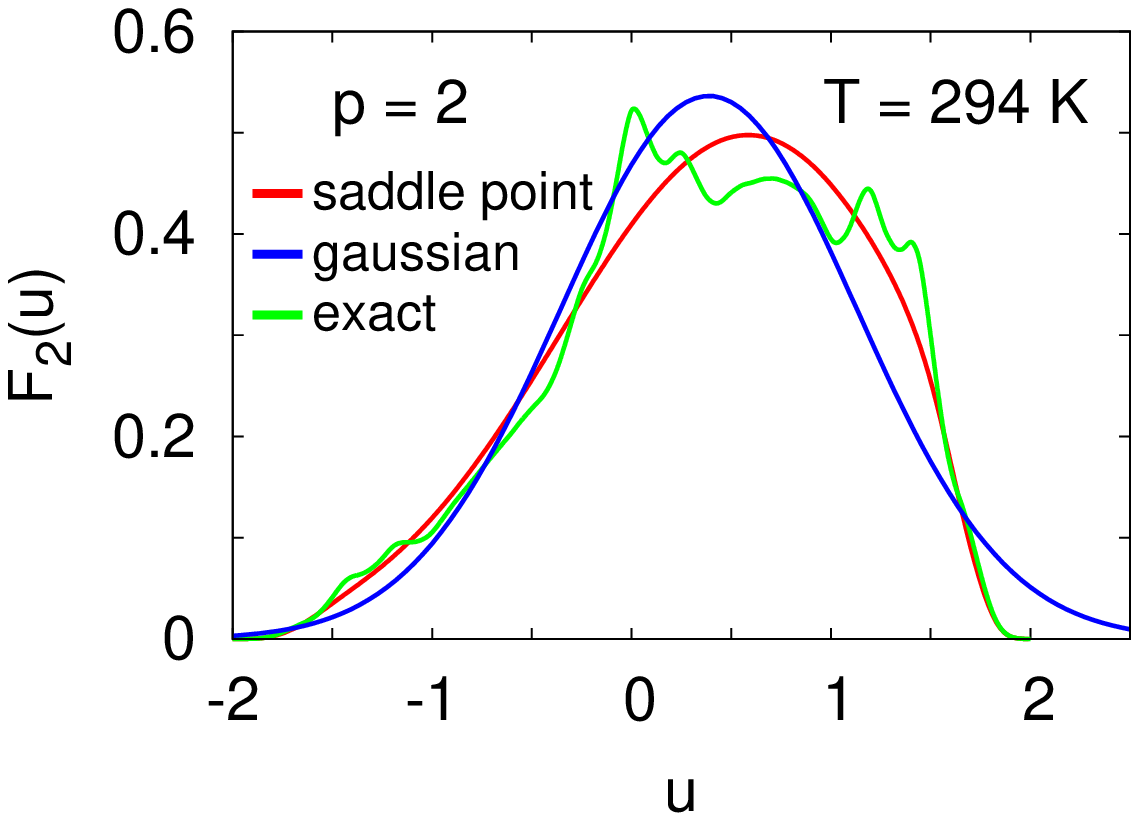}
\includegraphics[width=0.32\linewidth,angle=0]{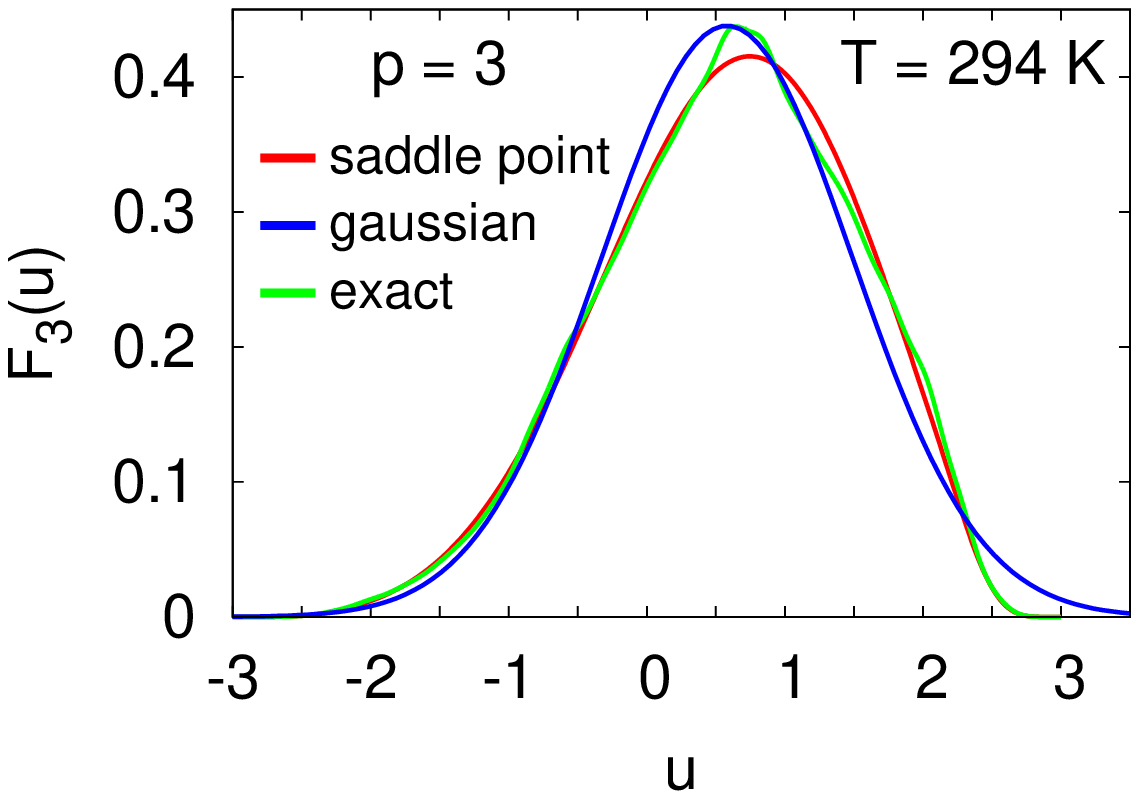}
\includegraphics[width=0.32\linewidth,angle=0]{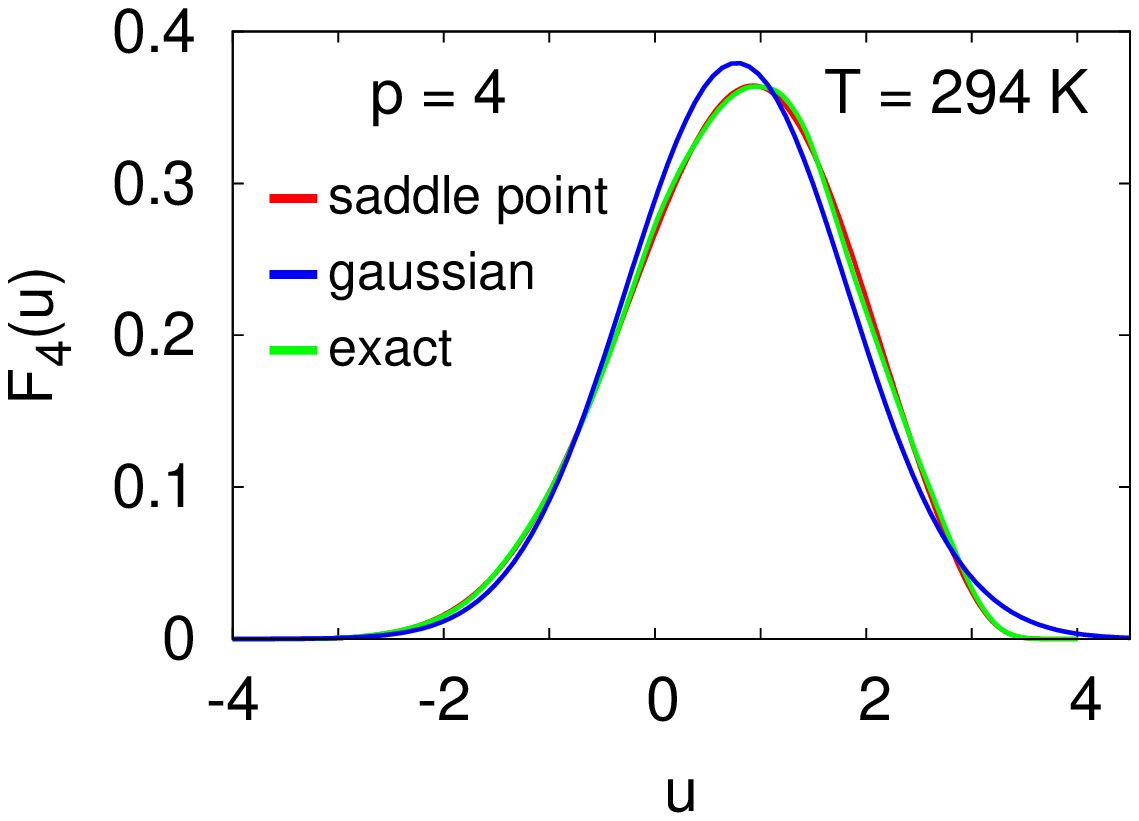}

\includegraphics[width=0.32\linewidth,angle=0]{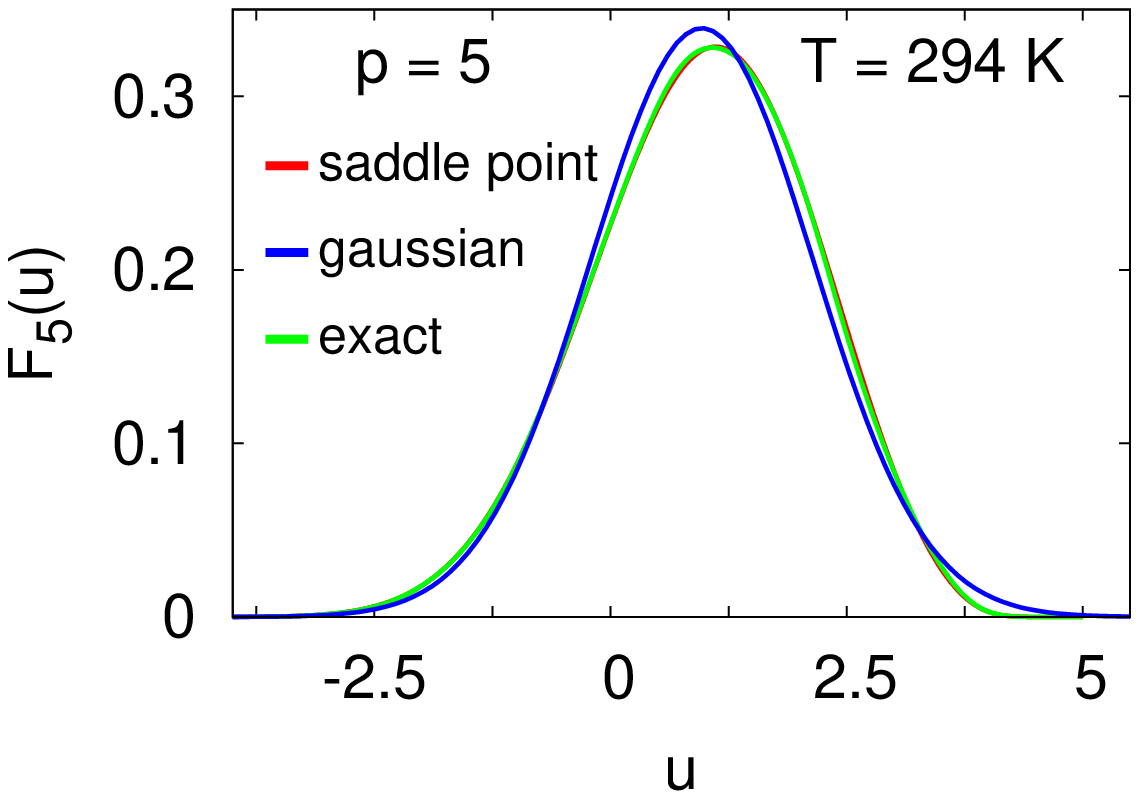}
\includegraphics[width=0.32\linewidth,angle=0]{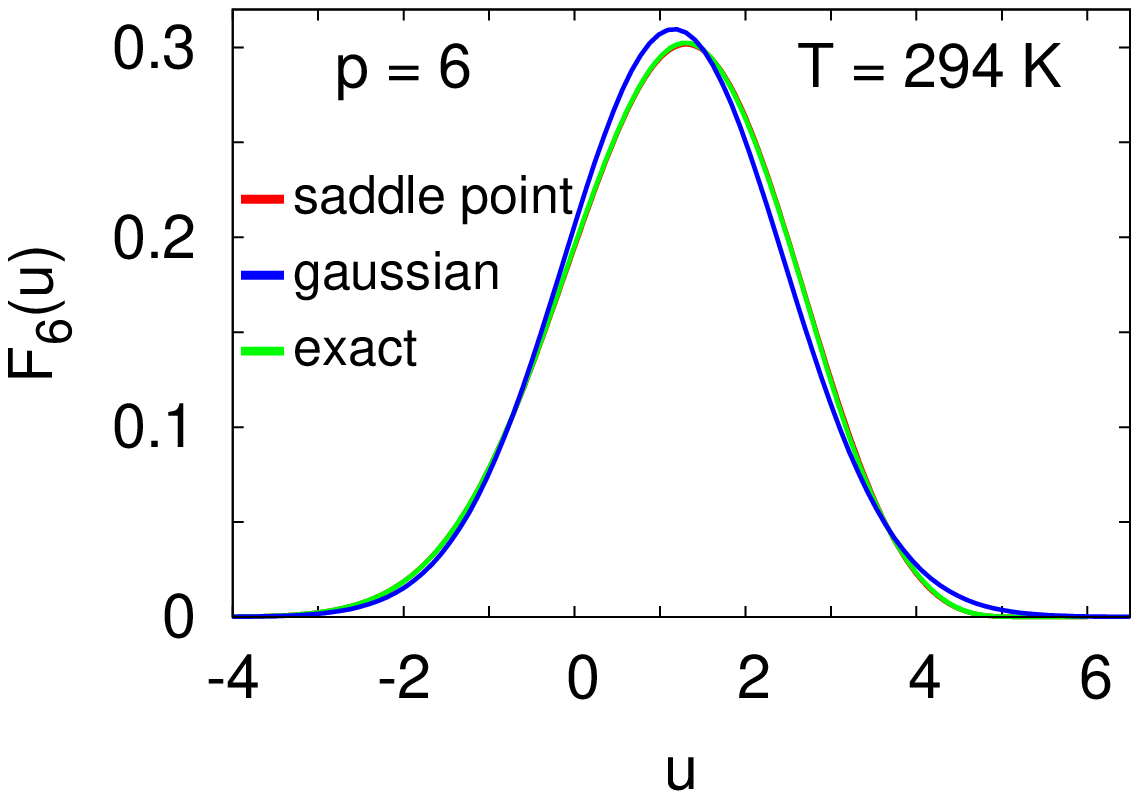}
\includegraphics[width=0.32\linewidth,angle=0]{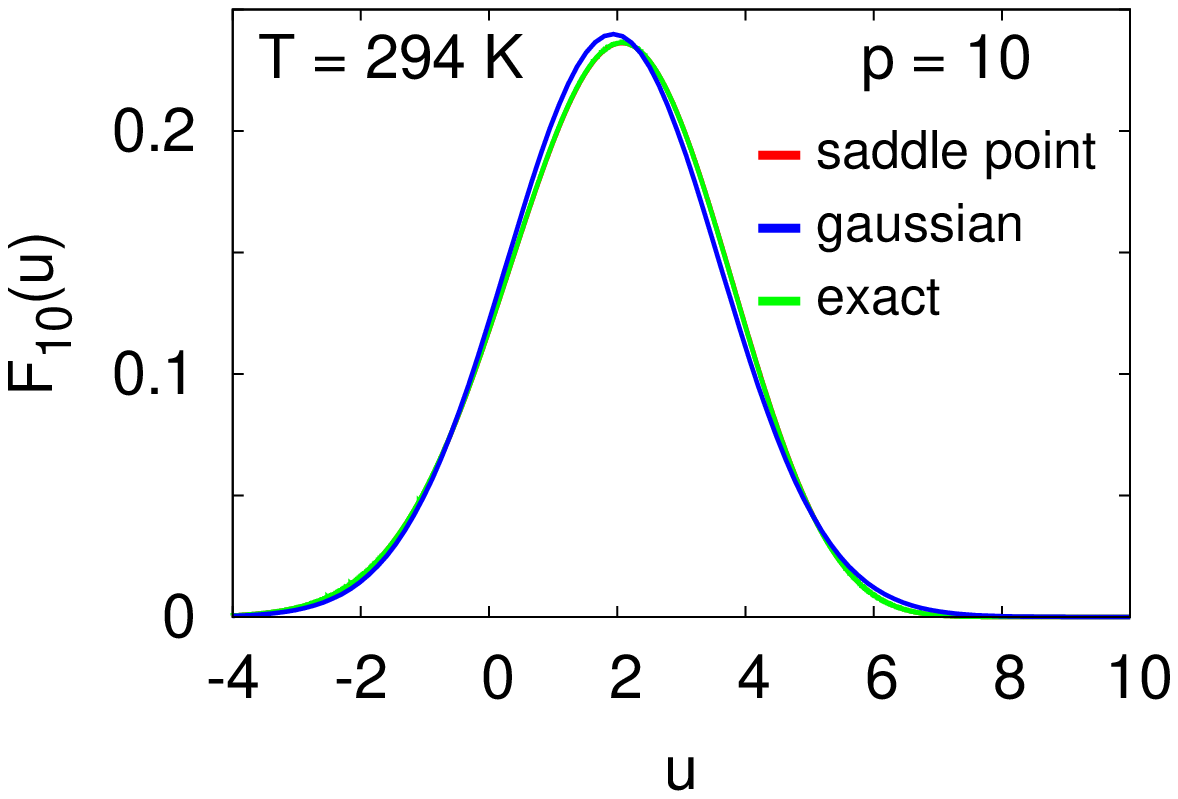}
\caption{The function $F_p(u)$ for vanadium at \mbox{294 K} computed in three ways: 
Eq.~(\ref{eq:Fpsp}), saddle point approximation (red);
Eq.~(\ref{eq:Fpg}), gaussian approximation (blue); 
and exact, conmputed by numerical integration (green).
\label{fig:Fps294}}
\end{figure}

\section{Numerical comparisons}
\label{sec:num}

To compare the different approximations for the scattering function we used vanadium, with the measured 
DoS published by Sears \textit{et al.}~\cite{Sears95}.

We compare the SPPE developed in this paper with the widely used gaussian approximation. 
As a control, the exact $F_p(u)$ is computed by numerical integration using 
the recursion relation~(\ref{eq:rr}). 
The results for $p$ from 2 to 6 and $p=10$ or $p=20$ are displayed in Figs.~\ref{fig:Fps77} 
and~\ref{fig:Fps294} for \mbox{77 K} and \mbox{294 K}, respectively. Notice that the SPPE
approximation is far more accurate than the gaussian approximation. Indeed, the saddle point
approximation is nearly exact for $p>2$, and is not bad for $p=2$.
The gaussian approximation is better for higher T and, as expected, improves by increasing $p$.
It is very good for $p\geq 20$ at \mbox{77 K}, and for $p\geq 10$ at \mbox{294} K. 

Obviously, the differences between the gaussian and the SPPE approximations for the 
representation~(\ref{eq:SMPprac}) of the scattering function diminish by increasing $n_{\mathrm{ph}}$.
The accuracy of the scattering function representation for given $n_{\mathrm{ph}}$ depends strongly
on the momentum transfer $q$. At low $q$ the multi-phonon term $S_{\mathrm{MP}}$ is only important
if $n_{\mathrm{ph}}=1$. In that case, the SPPE
is much more accurate than the gaussian approximation, due to the poor representation provided by 
the latter for the two and three phonon terms. 
At high $q$ the main contribution comes from terms with large $p$, and the 
gaussian approximation is almost as good as the SPPE.
The SPFS, Eq.~(\ref{eq:Ssp}) is accurate only at very high values of momentum transfer.

In an intermediate range of $q$, however, the multi-phonon contribution is dominated
by terms with moderate $p$, not too low but not too high. In this case the SPPE
is notably more accurate than the gaussian approximation. 
The range of $q$ at which the different approximations are good depends strongly on the temperature.

\begin{figure}[t!]
\centering
\includegraphics[width=0.33\linewidth,angle=0]{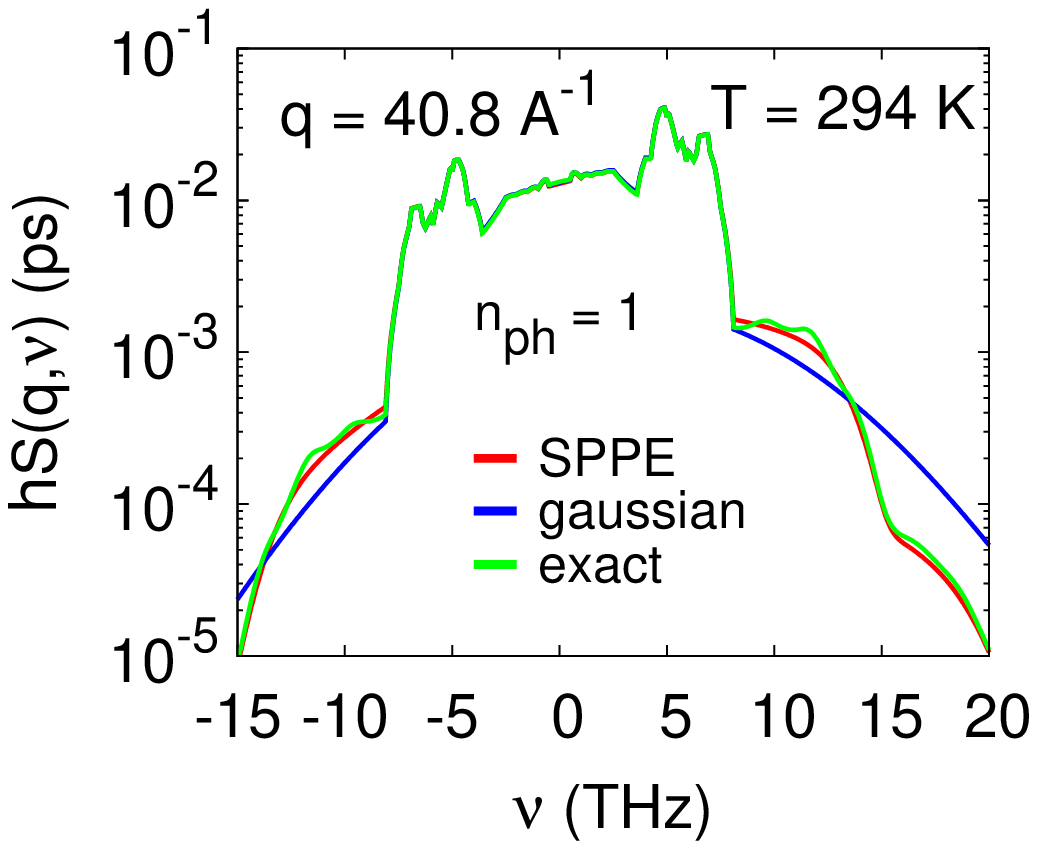}
\hspace{-0.7truecm}
\includegraphics[width=0.33\linewidth,angle=0]{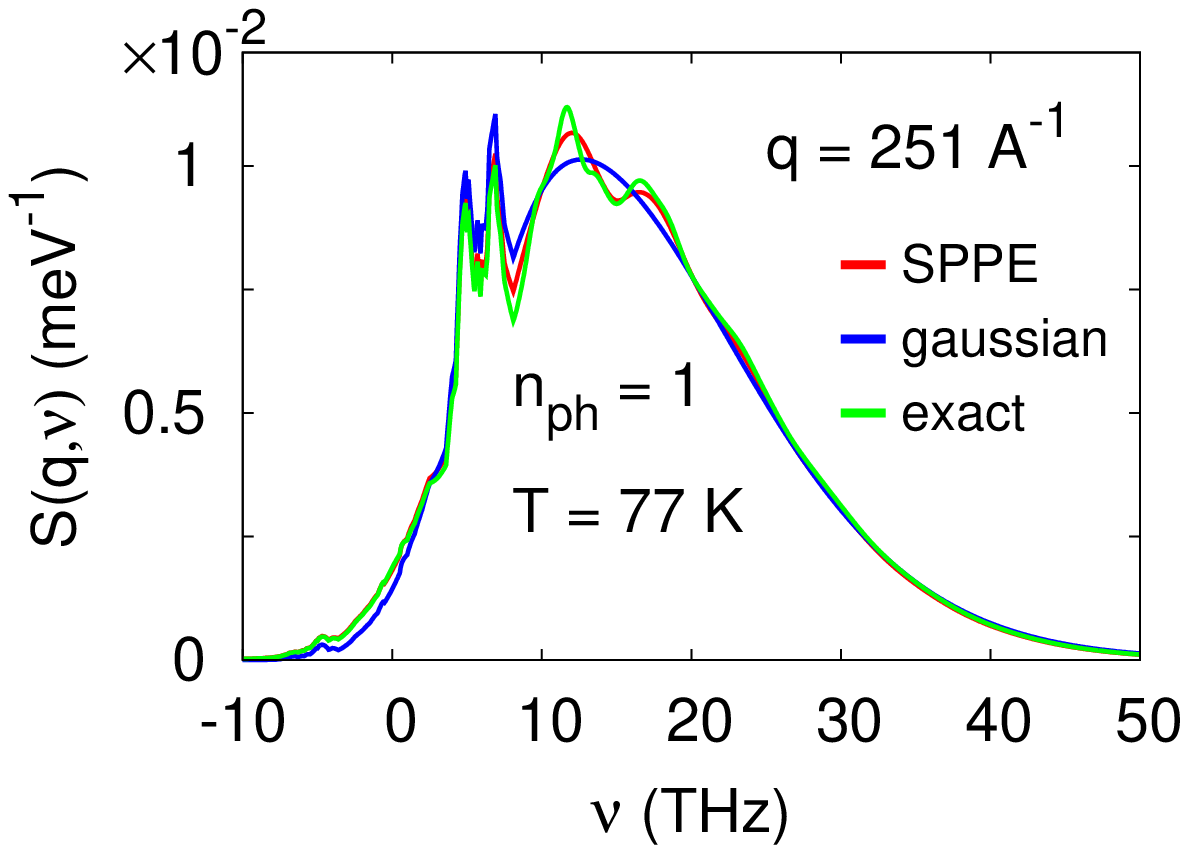}
\includegraphics[width=0.33\linewidth,angle=0]{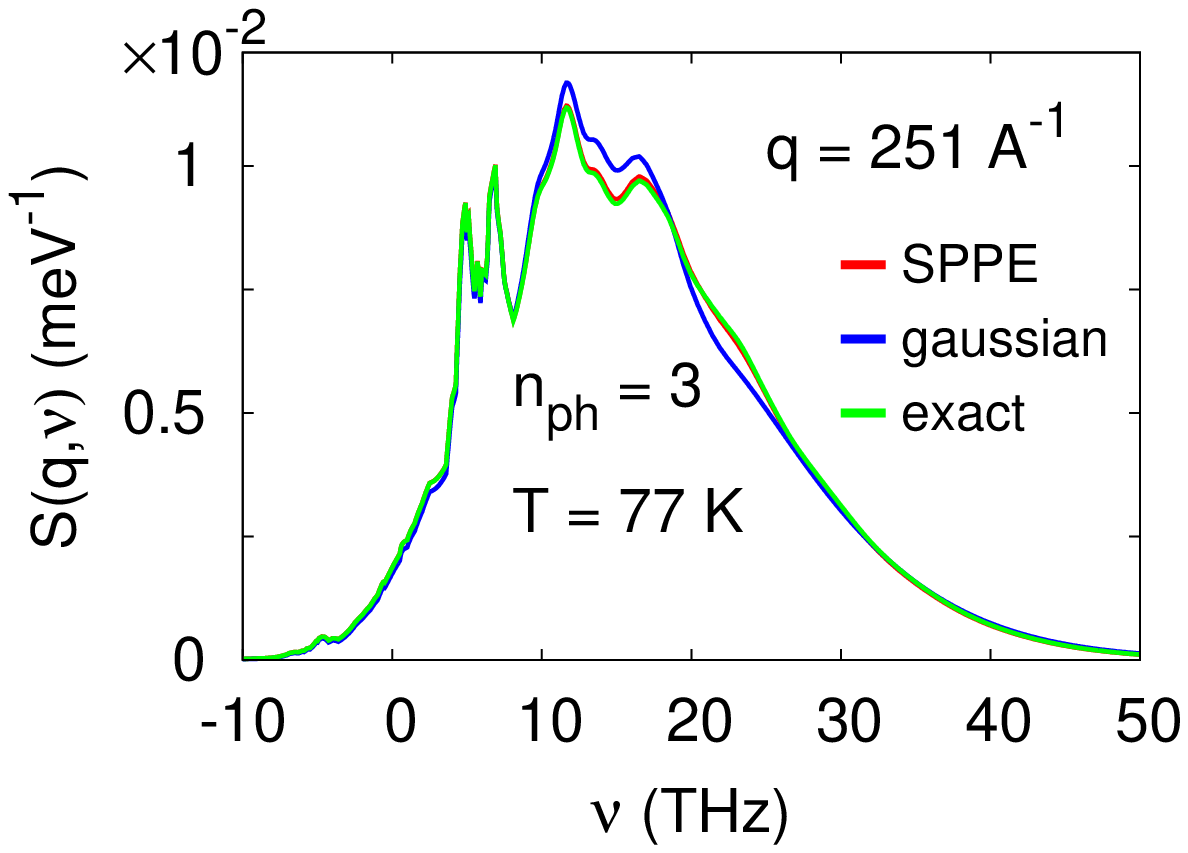}
\caption{The inelastic part of the scattering function of vanadium represented 
via Eq.~(\ref{eq:Sprac}) for several values of $n_{\mathrm{ph}}$, computed by 
the SPPE (red) and by the gaussian approximation (blue). 
The exact scattering function is plotted in green. 
\label{fig:S}}
\end{figure}

The discussion of the above paragraphs is illustrated in Fig.~\ref{fig:S}, where the inelastic part 
of the scattering function is displayed as a function of the frequency, $\nu$, in different situations.
The left panel shows $h S(q,\nu)$ at \mbox{294 K},
computed with $n_{\mathrm{p}}=1$. It can be directly compared with Fig.~9 of Ref.~\cite{Sears95}. 
Notice that the SPPE is very accurate, while the gaussian approximation
deviates notably from the exact results in the tails, due to its poor representation of the two 
and three phonon terms. The right panels correspond to \mbox{77 K} and computations with 
$n_{\mathrm{ph}}=1$ and 3. Notice that even with $n_{\mathrm{p}}=3$ the SPPE approximation is
noticeably more accurate than the gaussian approximation.

\begin{figure}[t!]
\centering
\includegraphics[width=0.32\linewidth,angle=0]{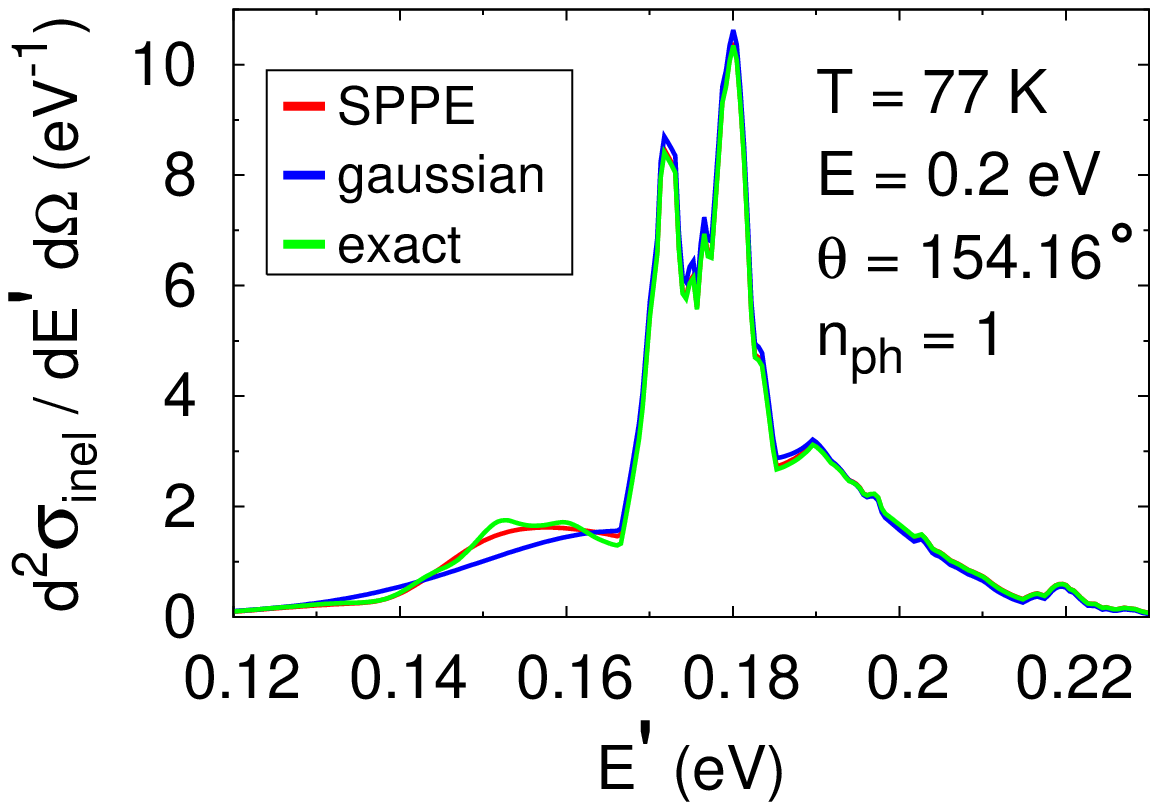}
\includegraphics[width=0.32\linewidth,angle=0]{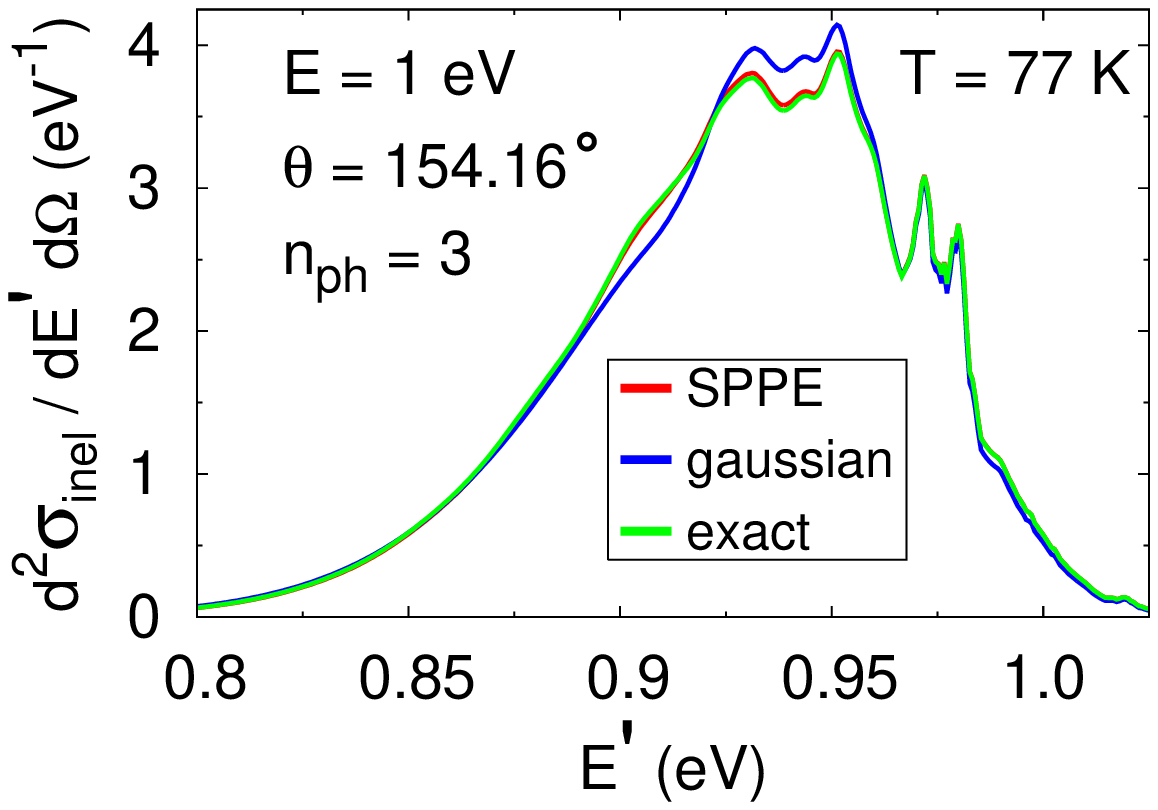}
\includegraphics[width=0.32\linewidth,angle=0]{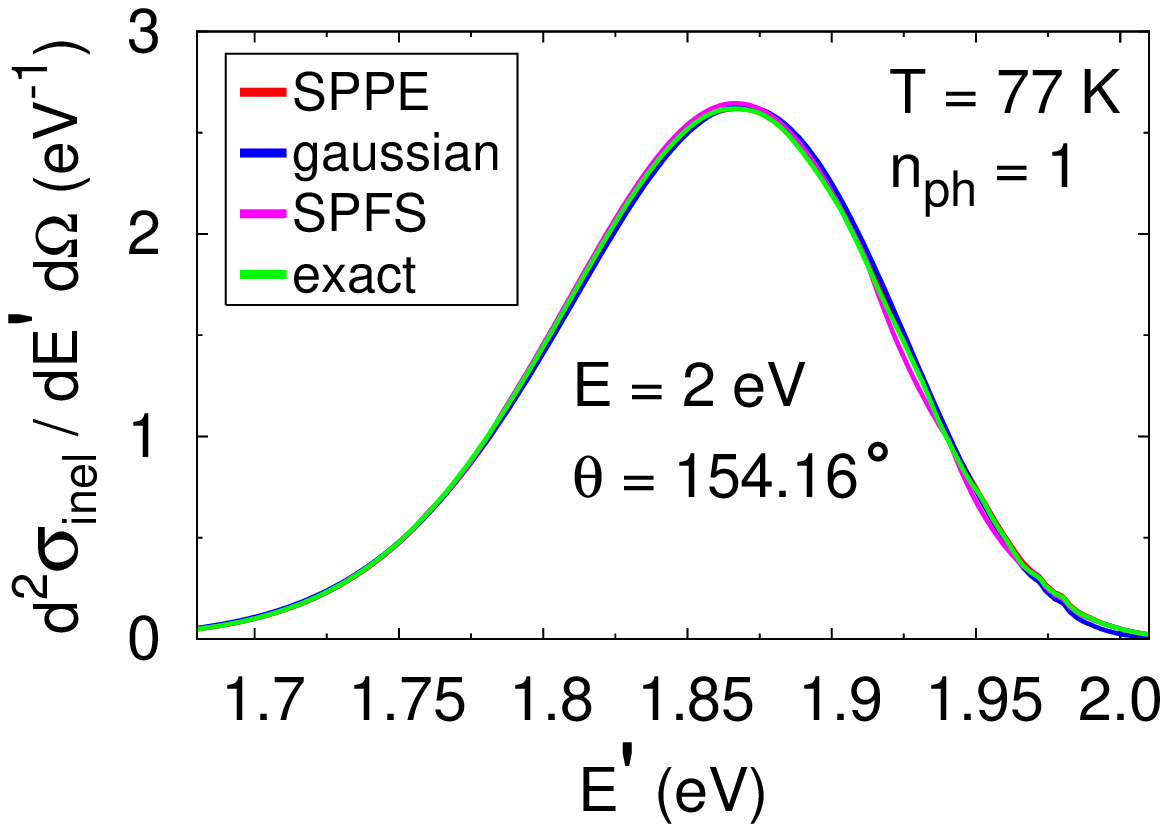}

\includegraphics[width=0.32\linewidth,angle=0]{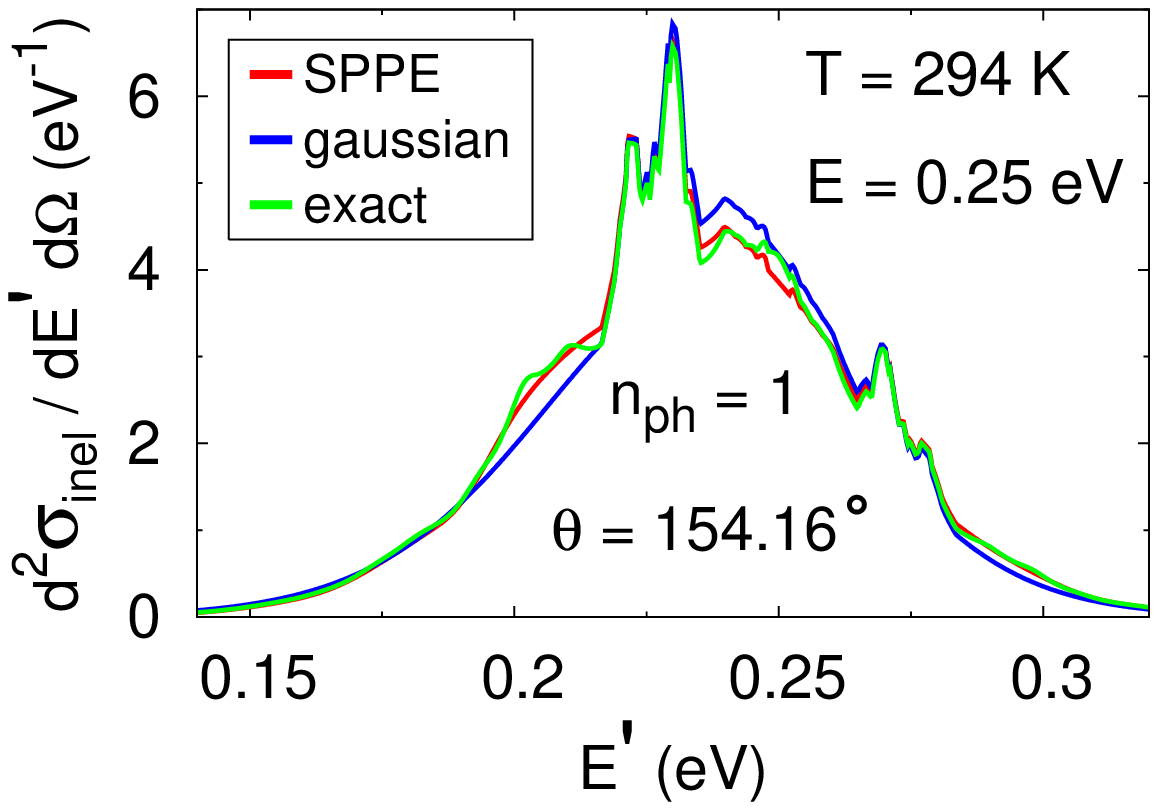}
\includegraphics[width=0.32\linewidth,angle=0]{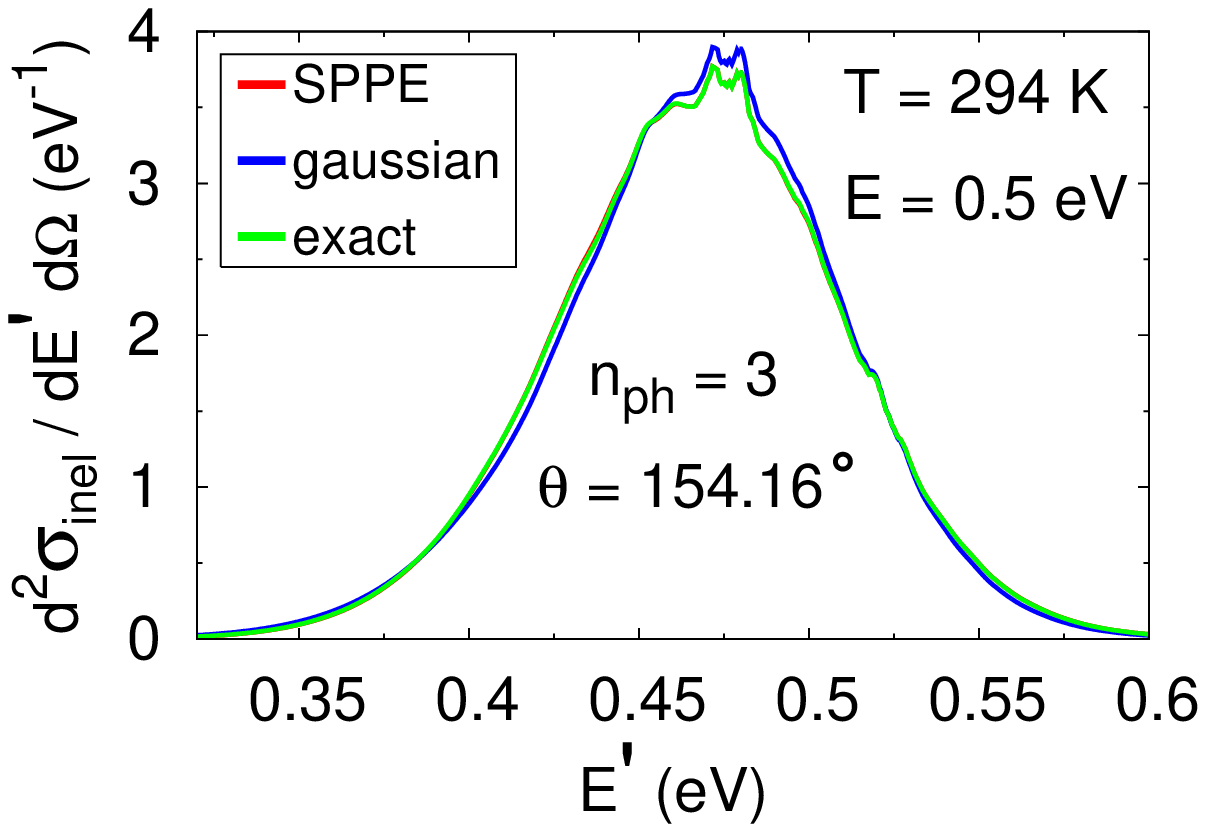}
\includegraphics[width=0.32\linewidth,angle=0]{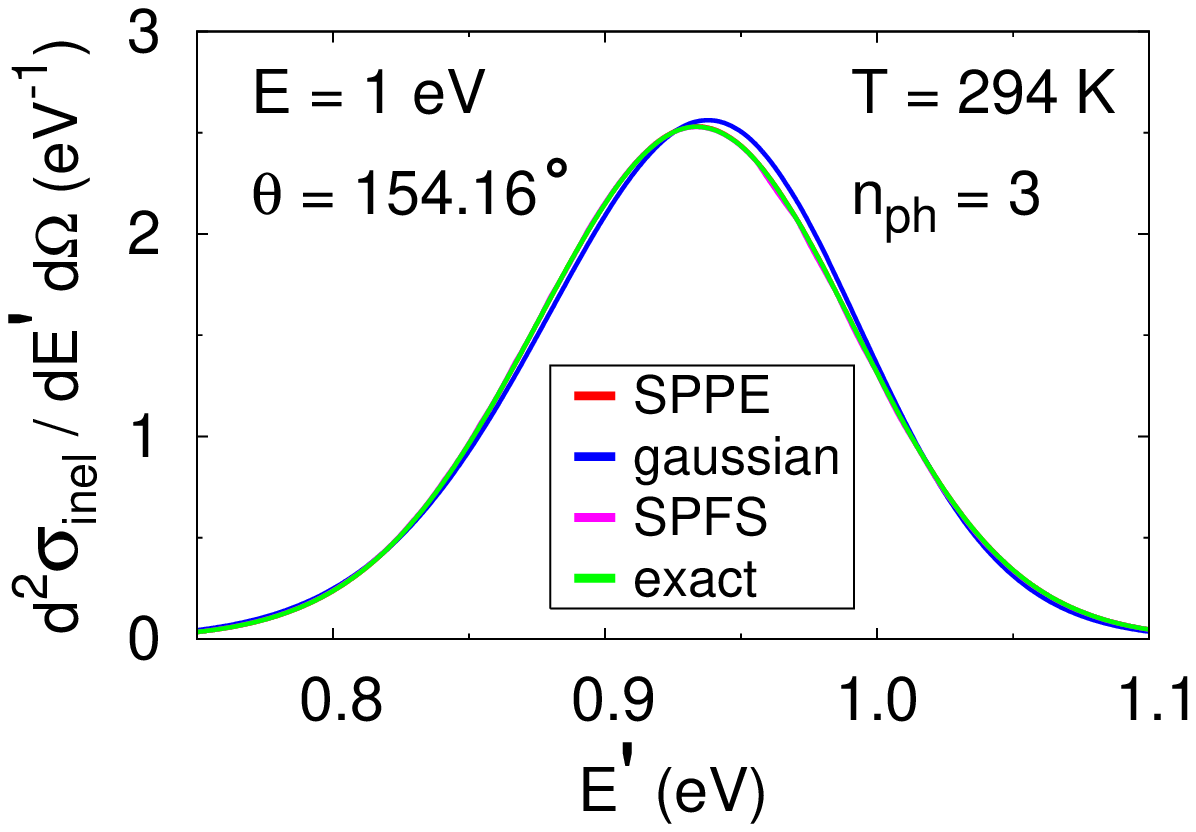}
\caption{The double differential inelastic scattering cross section for vanadium 
with the scattering function represented via Eq.~(\ref{eq:Sprac}),
computed with the SPPE (red), with the gaussian approximation (blue),
and with the SPFS (pink). The exact scattering cross section is plotted in green. 
\label{fig:XS}}
\end{figure}

Some examples of the double differential scattering cross section~(\ref{eq:XS}) computed with the
different approximations are displayed as a function of the scattered neutron energy, $E^\prime$,
for fixed values of the scattering angle, in Fig.~(\ref{fig:XS}). In all cases the scattering 
angle is $\theta=154.16^{\mathrm{o}}$. 
The top panels correspond to \mbox{77 K}.
In the left panel the incident neutron energy is $E=0.2$~eV, and $n_{\mathrm{ph}}=1$. The SPPE
approximation is very good but the gaussian approximation shows notably departures from the exact
result in an interval of the scattered neutron energy, $E^\prime$. But at this relatively low
incident energy the gaussian approximation is almost as good as the SPPE
if $n_{\mathrm{ph}}=2$, and both are nearly indistinguishable if $n_{\mathrm{ph}}=3$. This is obviously
due to the fact that multi-phonon terms with $p>3$ contribute very little to the cross section.
The middle panel displays the results for $E=1$~eV and $n_{\mathrm{ph}}=3$.
The SPPE approximation is extremely accurate, but the gaussian approximation 
is rather inaccurate, even though $n_{\mathrm{ph}}=3$. 
The right panel corresponds to a relatively high energy, $E=2$~eV. The momentum transfer
is high enough that the SPFS is very good.
Thus, in this case the three approximations considered in this paper are very accurate,
although the SPPE is still the more accurate. 
A completely similar discussion can be made for the bottom panels, that correspnd to \mbox{294 K}.

\section{Conclusions}
\label{sec:conc}

The saddle point approximation for the phonon expansion, Eq.~(\ref{eq:phe}), is more accurate than 
the well known gaussian approximation developed in Refs. \cite{Sjolander58}, \cite{Schofield59}, 
and \cite{Lovesey84}.
The differences are more important for the lowest order terms of the phonon expansion, for which
the gaussian approximation shows important departures from the exact result, while the SPPE
approximation is very good even for $p=2$ and nearly exact for $p>2$.
The gaussian approximation improves by increasing $p$ and becomes very accurate, basically 
indistinguishable from the SPPE, 
if $p$ is large enough. The reason is that the gaussian approximation is based on an expansion 
of $\gamma(s)$ around $s=0$, and the complex saddle point $\ui\tph$ tends to zero as 
$p\rightarrow\infty$.

Nevertheless, the gaussian approximation gives generally very good results and
has the virtue of being highly simple, depending only on two parameters, $\gamma_0$ and
$\gamma^{\,\prime\prime}(0)$, that can be computed as integrals involving the DoS.
On the other hand, the SPPE requires
three functions, $\tph(\xi)$, $\gamma(\ui\tph)$, and $\gamma^{\,\prime\prime}(\ui\tph)$.
The function $\tph(\xi)$ is obtained by solving the saddle point equation. 
Although this is more complicated than the simpler gaussian approximation, it is still
simple enough to be used in practice, for instance in Monte Carlo simulations.
The three functions can be precomputed and stored in tables read by the Monte Carlo program,
that used them to compute the scattering cross sections by interpolation.
This will improve the accuracy of the results in cases of interest.

\vspace{0.5truecm}

\noindent
\textsc{Acknowledgements} \\
This project has received funding from the European Union\textsc{\char13}s 
Horizon 2020 research and innovation programme under grant agreement No 654000.
The authors acknowledge the Grant No. MAT2015-68200- C2-2-P from the Spanish
Ministry of Economy and Competitiveness. This work was
partially supported by the scientific JSPS Grant-in-Aid
for Scientific Research (S) (Grant No. 25220803), and the
MEXT program for promoting the enhancement of research
universities, and JSPS Core-to-Core Program, A. Advanced
Research Networks.


\appendix

\section*{Appendix}

In this appendix some interesting features on the saddle point equation~(\ref{eq:spph}) and 
its solution are studied.

\subsection*{Saddle point equation}

Let us start by writing the saddle point equation in a suggestive form. First, notice that the integrand defining 
$\gamma(s)$ in Eq.~(\ref{eq:gdef}) is positive for $s=\ui\tph $. Thus, it can be interpreted as a
probability distribution for $u$, with compact support in the interval $[-1,1]$. The expectation value of any
function $f(u)$ with this probability distribution is given by
\begin{equation}
\langle f(u)\rangle_{\tph } = \frac{1}{\gamma(\ui\tph )}
\int_{-\infty}^\infty du^\prime f(u^\prime) \frac{Z(u^\prime)}{u^\prime} n(u^\prime\omega_{\mathrm{m}}) \mathrm{e}^{\tph u^\prime}.
\label{eq:evf}
\end{equation}
Thus, the saddle point equation~(\ref{eq:spph}) can be written as
\begin{equation}
\langle u\rangle_{\tph } = -\xi.
\label{eq:spph2}
\end{equation}
Since $-1<\langle u\rangle_{\tph }<1$, the saddle point equation has no solution if $|\xi|>1$,
\textit{i.e.} if $|u|>p$.
It is also clear from the form of $\gamma(s)$ that 
\begin{equation}
\lim_{\tph\rightarrow\pm\infty} \langle u\rangle_{\tph } = \mp 1.
\end{equation}
Conversely, the solution of the saddle point equation, $\tph (\xi)$, diverges to $\mp\infty$ in the
limit $\xi\rightarrow\pm 1$. 

Notice also the relation
\begin{equation}
-\frac{\gamma^{\,\prime\prime}(\ui\tph)}{\gamma(\ui\tph)}-\frac{u^2}{p^2} = 
\langle u^2\rangle_\tph - \langle u\rangle_\tph^2 > 0.
\end{equation}
Similarly, notice that $\Delta^2$ of Eq~(\ref{eq:Delta2}) can be written as 
$\Delta^2=\langle u^2\rangle_0 - \langle u\rangle_0^2$ and it is thus positive.

\subsection*{Asymptotic behavior of $\tph(\xi)$ for $\xi\rightarrow\pm 1$}

The asymptotic form of $\tph (\xi)$ in such limits has the generic form
\begin{equation}
\tph = \mp\frac{B_{\mathrm{s}}}{1\mp\xi} \mp B_0 + \ldots
\label{eq:tph_as}
\end{equation}
where the coefficients $B_{\mathrm{s}}$, $B_1$, ... depend on temperature and on the behavior of 
$Z(u)$ as $u\rightarrow 1$. Let us assume that this behavior is
\begin{equation}
Z(u) = Z_\mathrm{s} (1-u)^\alpha\left[1 + Z_1(1-u) + Z_2(1-u)^2 + \ldots \right],
\label{eq:Z1}
\end{equation}
were $\alpha\geq 0$ and $Z_1,\ldots$ are coefficients. Notice that for a Debye model $\alpha=0$.

To get the asymptotic behavior of $\tph(\xi)$ as $\xi\rightarrow\pm 1$
we make the change of variable $u^\prime=\mp 1 - x/\tph $ in the integrals entering (\ref{eq:evf}).
One has to bear in mind that $\tph\rightarrow \mp\infty$. Expanding the integrands in
powers of $x/\tph$, using Eq.~(\ref{eq:Z1}), we get an asymptotic series in powers
of $1/\tph$ for $\langle f(u)\rangle_\tph$. Setting $f(u)=u$ we get 
\begin{equation}
\langle u\rangle_{\tph} = \mp \left(1  - \frac{C_1}{\tph } - \frac{C_2}{\tph^2} + \ldots\right),
\label{eq:evu_as}
\end{equation}
with
\begin{eqnarray}
C_1 &=& \alpha+1, \\
C_2 &=& 1 + \frac{\hbar\omega_\mathrm{m}}{k_\mathrm{B}T}n(\omega_\mathrm{m}) + Z_1.
\end{eqnarray}
Plugging Eq.~(\ref{eq:evu_as}) into~(\ref{eq:spph2}) and solving for $\tph$ we obtain Eq.~(\ref{eq:tph_as})
with $B_\mathrm{s}=C_1$ and $B_0=C_2/C_1$.

\subsection*{Form of $F_p^{\mathrm{(sp)}}(u)$ for $u\rightarrow\pm p$}

To get the form of $F_p(u)$ for $u\rightarrow \pm p$ we plug Eq.~(\ref{eq:tph_as}) into Eq.~(\ref{eq:Fpsp}),
with $\xi=\pm u/p$, and expand in powers of $1\mp u/p$, obtaining
\begin{equation}
F_p^{\mathrm{(sp)}}(u) = \sqrt{\frac{\alpha+1}{2\pi p}} 
\left(\mp\frac{Z_\mathrm{s}\Gamma(\alpha+1)\mathrm{e}^{\alpha+1}}{\gamma_0(\alpha+1)^{\alpha+1}}
n(\mp\omega_\mathrm{m})
\right)^p 
\left(1\mp\frac{u}{p}\right)^{p(\alpha+1)-1} + \ldots
\end{equation}
where $\Gamma(x)$ is the Euler Gamma function. Thus, $F_p(u)$ vanishes for $u\rightarrow\pm p^{\mp}$.

\subsection*{Asymptotic form of $F_p(u)$ for large $p$ around the maximum}

Notice first that from the saddle point equation we readily obtain
\begin{equation}
\frac{\partial\tph}{\partial\xi} = \left[\frac{\gamma^{\,\prime\prime}(\ui\tph)}{\gamma(\ui\tph)}
+\frac{u^2}{p^2}\right]^{-1}.
\end{equation}
Using this relation, we can easily get the derivative of $F_p^{\mathrm{(sp)}}(u)$ with respect to $u$.
From it, we have that the maximum $u_{\mathrm{m}}$ of this function satisfy the equation
\begin{equation}
\tph = \frac{\frac{\ui \gamma^{\,\prime\prime\prime}}{\gamma} 
- 3 \frac{\ui \gamma^{\,\prime}}{\gamma}\frac{\gamma^{\,\prime\prime}}{\gamma}
-2 \left(\frac{\ui \gamma^{\,\prime}}{\gamma}\right)^3}
{2p\left[\frac{\gamma^{\,\prime\prime}}{\gamma}+\frac{u^2}{p^2}\right]}
\end{equation}
where the argument of the functions, $\ui\tph$, has been omited for clarity.

For large $p\rightarrow\infty$ the maximum of $F_p^{\mathrm{(sp)}}(u)$ is attained
at $\tph\rightarrow 0$, and thus the gaussian approximation is obtained in this limit.
Plugging the above equation into the saddle point equation and solving to leading order
as $p\rightarrow\infty$ we have
\begin{equation}
u_{\mathrm{m}}=\frac{p}{\gamma_0}.
\end{equation}

\bibliographystyle{unsrt}
\bibliography{refs_neutrons}

\end{document}